\documentclass[apj]{emulateapj}

\usepackage{graphicx}

\shorttitle{Verifying the Cosmological Utility of SNe~Ia}
\shortauthors{Ellis {\it et al.}}

\begin{document}

\newcommand{\bq}{\begin{equation}}
\newcommand{\eq}{\end{equation}}

\newcommand\omatter{\ifmmode \Omega_{\mathrm{M}}\else $\Omega_{\mathrm{M}}$\fi}
\newcommand\ok{\ifmmode \Omega_{\mathrm{k}}\else $\Omega_{\mathrm{k}}$\fi}
\newcommand\olambda{\ifmmode \Omega_{\Lambda}\else $\Omega_{\Lambda}$\fi}
\newcommand\dmB{\ifmmode \Delta m_{15}(B) \else $\Delta m_{15}(B)$\fi}
\newcommand\zspec{\ifmmode z_{\mathrm{spec}}\else $z_{\mathrm{spec}}$\fi}
\newcommand\zphot{\ifmmode z_{\mathrm{phot}}\else $z_{\mathrm{phot}}$\fi}
\newcommand\ebmvmw{\ifmmode E_{B-V}^{\small \mathrm{mw}}\else $E_{B-V}^{\small \mathrm{mw}}$\fi}
\newcommand\ebmvhost{\ifmmode E_{B-V}^{\small \mathrm{host}}\else $E_{B-V}^{\small \mathrm{host}}$\fi}
\newcommand\aperpix{\ifmmode \mathrm{\AA}\,\mathrm{pix}^{-1}\else \AA\,$\mathrm{pix}^{-1}$\fi}

\title{Verifying the Cosmological Utility of Type Ia Supernovae: 
Implications of a Dispersion in the Ultraviolet Spectra}

\author{
  R.~S.~Ellis\altaffilmark{1}, 
  M.~Sullivan\altaffilmark{2,3},
  P.~E.~Nugent\altaffilmark{4}, 
  D.~A.~Howell\altaffilmark{2},
  A.~Gal-Yam\altaffilmark{1}, 
  P.~Astier\altaffilmark{5}, 
  D.~Balam\altaffilmark{6}, 
  C.~Balland\altaffilmark{5}, 
  S.~Basa\altaffilmark{7}, 
  R.~G.~Carlberg\altaffilmark{2}, 
  A.~Conley\altaffilmark{2}, 
  D.~Fouchez\altaffilmark{8}, 
 J.~Guy\altaffilmark{5}, 
 D.~Hardin\altaffilmark{5}, 
 I.~Hook\altaffilmark{3}, 
 R.~Pain\altaffilmark{5}, 
  K.~Perrett\altaffilmark{2}, 
  C.~J.~Pritchet\altaffilmark{6}, 
 N.~Regnault\altaffilmark{5}
} 

\altaffiltext{1}{California Institute of Technology, E. California Blvd, Pasadena CA 91125, USA}
\altaffiltext{2}{Department of Physics and Astronomy, University of Toronto, 50 St. George Street, Toronto, ON M5S 3H4, Canada} 
\altaffiltext{3}{Department of Physics (Astrophysics), University of Oxford, Keble Road, Oxford OX1 3RH, UK}
\altaffiltext{4}{Lawrence Berkeley National Laboratory, 1 Cyclotron Road, Berkeley, CA 94720, USA}
\altaffiltext{5}{LPHNE, CNRS-IN2P3 and Universit\'{e}s Paris VI \& VII, 4 Place Jussieu, 75252 Paris Cedex 05, France}
\altaffiltext{6}{Department of Physics and Astronomy, University of Victoria, PO Box 2055 STN CSC, Victoria BC V8T1M8, Canada}
\altaffiltext{7}{LAM, CNRS, BP8, Traverse du Siphon, 13376 Marseille Cedex 12, France}
\altaffiltext{8}{CPPM, CNRS-IN2P3 and Universit\'{e} Aix-Marseille II, Case 907, 13288 Marseille Cedex 9, France}

\email{rse@astro.caltech.edu}

\begin{abstract}
  
We analyze the mean rest-frame ultraviolet (UV) spectrum of Type Ia
Supernovae (SNe) and its dispersion using high signal-to-noise
Keck-I/LRIS-B spectroscopy for a sample of 36 events at intermediate
redshift ($\overline{z}$=0.5) discovered by the Canada-France-Hawaii
Telescope Supernova Legacy Survey (SNLS).  We introduce a new method
for removing host galaxy contamination in our spectra, exploiting
the comprehensive photometric coverage of the SNLS SNe and their
host galaxies, thereby providing the first quantitative view of the UV
spectral properties of a large sample of distant SNe~Ia.  Although
the mean SN~Ia spectrum has not evolved significantly
over the past 40\% of cosmic history, precise evolutionary
constraints are limited by the absence of a comparable sample of
high quality local spectra.  The mean UV spectrum of our
$z\simeq$0.5 SNe~Ia and its dispersion is tabulated for use in
future applications. Within the high-redshift sample, we discover
significant UV spectral variations and exclude dust extinction as
the primary cause by examining trends with the optical SN color.
Although progenitor metallicity may drive some of these trends, the
variations we see are much larger than predicted in recent models
and do not follow expected patterns.  An interesting new result is a
variation seen in the wavelength of selected UV features with phase. 
We also demonstrate systematic differences in the SN~Ia spectral features
with SN lightcurve width in both the UV and the optical.  We show
that these intrinsic variations could represent a statistical
limitation in the future use of high-redshift SNe~Ia for precision
cosmology.  We conclude that further detailed studies are needed, both
locally and at moderate redshift where the rest-frame UV can be
studied precisely, in order that future missions can confidently be
planned to fully exploit SNe~Ia as cosmological probes.

\end{abstract}


\section{Introduction}
\label{sec:introduction}

Supernovae of Type Ia (SNe~Ia) are now well-established as
cosmological distance indicators. In addition to the original surveys
by the Supernova Cosmology Project
\citep[SCP;][]{1997ApJ...483..565P,1999ApJ...517..565P} and the High-Z
Supernova Search Team \citep{1998ApJ...507...46S,1998AJ....116.1009R},
a new generation of SN~Ia surveys is underway both locally
\citep{2002SPIE.4836...61A,2005coex.conf..525L,2006PASP..118....2H}
and at higher redshifts
\citep{2006A&A...447...31A,2007ApJ...659...98R,2007ApJ...666..694W}.
Despite the availability of independent probes of the presence and
properties of dark energy from studies of the cosmic microwave
background \citep{2007ApJS..170..377S} and galaxy redshift surveys
\citep{2002MNRAS.330L..29E,2005MNRAS.362..505C,2005ApJ...633..560E},
the luminosity distance--redshift relation for SNe~Ia provides the
only {\it direct} evidence for a cosmic acceleration.

The detection and characterization of dark energy, via measurements of
the average cosmic equation of state parameter $<$$w$$>$, requires the
precision measurement of SNe~Ia to redshifts $z\simeq$0.5--1, sampling
the epoch of cosmic acceleration \citep{2006A&A...447...31A}. However,
more precise constraints on the nature of dark energy, for example
evidence for any variation in $w$ with redshift, requires extending
these studies to redshift $z$$>$1
\citep{2004ApJ...607..665R,2007ApJ...659...98R} where the early
effects of deceleration may be detectable. As projects are developed
which plan to probe SNe~Ia beyond $z$=1 for this purpose
\citep[e.g][]{2005NewAR..49..346A,2006SPIE.6265E..67B}, it becomes
important to understand the possible limitations of using SNe~Ia as
distance probes.  Key issues relating to the diversity of SNe~Ia and
their possible evolution with redshift as a population, together with
the limiting effects of dust and/or color corrections to their
photometric properties, are particularly crucial to understand.

Several local studies \citep{1995AJ....109....1H,1999AJ....117..707R,
  2000AJ....120.1479H,2001ApJ...554L.193H,
  2005A&A...433..807M,2005ApJ...634..210G} have already indicated
correlations between SN~Ia properties and host galaxy morphologies.
More recently, \citet{2006ApJ...648..868S} have shown that the
properties of distant SNe~Ia appear to be a direct function of their
local stellar population, with the distribution of light curve widths
and hence peak luminosities correlating with the host galaxy specific
star-formation rate.  This work also determined that the rate of
SNe~Ia per unit stellar mass of their host galaxies is larger in
actively star-forming galaxies, suggesting many must be produced quite
rapidly in recently formed stellar populations, perhaps suggestive of
more than one progenitor mechanism. The authors conclude that SNe~Ia
may well be a bimodal or a more complex population of events
\citep[see also][]{2005ApJ...629L..85S,2006MNRAS.370..773M}.  Such
diversity in the properties of SNe~Ia could have far-reaching
implications, particularly if the {\it mix} of mechanisms or delay
times within the broad population gradually changes with look-back
time
\citep[e.g.][]{2006MNRAS.370..773M,2006ApJ...648..868S,2007astro.ph..1912H}.

These recent developments, which illustrate how improved precision
reveals new physical correlations in the SN~Ia population, raise the
broader question of whether future SN~Ia experiments might be limited
in precision by variations of a systematic nature within the
population, for example with redshift, which cannot be removed via
empirical correlations.  Detailed local surveys such as the LOSS/KAIT
\citep{2005coex.conf..525L} and CfA surveys
\citep{1999AJ....117..707R,2006AJ....131..527J} have presented
valuable data on the homogeneity and trends in the SNe~Ia population.
Further promising work is being undertaken via the Supernova Factory
\citep{2002SPIE.4836...61A} and the Carnegie Supernova Project
\citep{2006PASP..118....2H}. Important though these continued programs
will be, they are insufficient to address all possible concerns about
the use of SNe~Ia as precision tools in cosmology.  Comparable studies
at intermediate redshift\footnote{Defined here to represent the range
  0.2$<$$z$$<$0.7.} will be particularly important in order to address
questions relating to possible evolutionary effects and environmental
dependencies. In addition, it is not always practical at low redshift
to cover the full wavelength range necessary to test for systematic
trends.

In this paper we analyze high signal-to-noise ratio rest-frame
ultraviolet (UV) spectra of a large sample of intermediate redshift
SNe~Ia drawn from the Canada-France-Hawaii Telescope Supernova Legacy
Survey \citep[SNLS;][]{2006A&A...447...31A}, a rolling search which is
particularly effective for locating and studying events prior to their
maximum light. Our aim is obtain a substantially higher
signal-to-noise in the spectra than that typically obtained during
spectroscopic programs to type SNe and measure redshifts. We target
the UV wavelength region because in this wavelength region the SN
spectrum is thought to provide the most sensitive probe of {\it
  progenitor metallicity}
\citep[e.g.][]{1998ApJ...495..617H,2000ApJ...530..966L}, a variable
which may shed light on the possibility of progenitor evolution. The
time-dependent UV spectrum is also needed in estimating ``cross-band''
$k$-corrections, particularly at redshifts $z>1$ where optical
bandpasses probe the rest-frame near-UV
\citep{2004ApJ...607..665R,2007ApJ...659...98R}. Little is known about
the properties and homogeneity of the UV spectra of SNe~Ia, largely
because of the absence of suitable instruments for studying this
wavelength range in local events. Although some local SN~Ia UV spectra
are available from International Ultraviolet Explorer (IUE) or Hubble
Space Telescope (HST) satellite data
\citep[e.g.][]{1991ApJ...371L..23L,1993ApJ...415..589K,1995ESASP1189.....C},
the bulk of the progress now possible in this area can be provided
from optical studies of intermediate-redshift events with large
ground-based telescopes.

The goals of this paper are thus to address the question of both the
diversity and possible physical evolution in the intermediate redshift
SN~Ia family. We compare the rest-frame UV behavior of local SNe~Ia
with that derived for intermediate-redshift ($z\simeq$0.5) events
where the rest-frame UV enters the region of high efficiency of the
Keck LRIS-B spectrograph.  We also study the degree to which the UV
spectra at intermediate redshift represent a homogeneous population,
independent of other variables such as the physical environment and
light curve stretch.

A plan of the paper follows. In $\S$~\ref{sec:selection-cfhtls-sne} we
introduce the salient features of the SNLS and our method for
selecting SNe~Ia for detailed study. In
$\S$~\ref{sec:keck-observations} we discuss the Keck spectroscopic
observations and their reduction, including the treatment of host
galaxy subtraction and flux calibration. In $\S$~\ref{sec:analyses} we
consider our sample with respect to the broader set of SNe found by
SNLS, ensuring it is a representative subset in terms of various
observables, and discuss existing local UV spectra. In
$\S$~\ref{sec:results}, we undertake the detailed analysis. First we
compare the UV spectra found in our sample with those found locally.
We then examine the diversity of intermediate-redshift SNe~Ia in
various ways and correlate the UV variations with the light curves of
the SNe and the properties of the host galaxies. We discuss these
trends in terms of progenitor mechanisms in $\S$~\ref{sec:discussion}
and examine the implications in terms of possible long term
limitations of SNe~Ia as probes of dark energy.  We also present the
mean phase-dependent SN Ia spectrum and its uncertainties for use in
future work.

\section{Selection of SNLS SNe~Ia}
\label{sec:selection-cfhtls-sne}

Our SNe~Ia are taken from the Supernova Legacy Survey (SNLS), a
``rolling'' search for distant SNe with a primary science goal of
determining the average equation-of-state parameter of dark energy,
$<$$w$$>$ \citep[see][]{2006A&A...447...31A}. SNLS exploits the
square-degree Megacam camera \citep{2003SPIE.4841...72B} on the
Canada-France-Hawaii Telescope (CFHT), and comprises repeat imaging in
4 filters, $g'r'i'z'$, of four deep 1$\times$1 degree fields (see
\citeauthor{2006AJ....131..960S} \citeyear{2006AJ....131..960S} for
the field coordinates), plus further $u^{\star}$ imaging which is not
time-sequenced. Each field is imaged several ($\sim5$) times per
lunation for 5-6 lunations per year. A description of the real-time
search operations and the criteria for following SN candidates
spectroscopically can be found in \citet{2006AJ....131..960S}.
Spectroscopic follow-up time for the essential work of basic redshift
measurement and SN type determination for cosmological analyses comes
from major long-term programs at the European Southern Observatory
Very Large Telescopes (Basa et al., in prep; PI: Pain), the Keck observatory
(PI: Perlmutter) and the Gemini North and South telescopes \citep[][PI:
Hook]{2005ApJ...634.1190H,2007bronder}.

Observing time with the Keck-I telescope for the detailed
spectroscopic study of individual SNe presented in this paper was
scheduled from 2003 through 2005. The goal was to obtain substantially
higher signal-to-noise ratio spectra than those required for SN typing
(particularly in the rest-frame UV), and hence we targeted SNe~Ia with
a mean redshift lower than that of the SNLS as a whole
($\overline{z}$=0.45 versus $\overline{z}$=0.6) and used integration
times $\sim$ 3--4 times longer for that redshift. 

To avoid spectroscopic screening of every candidate (results from the
mainstream VLT and Gemini programs were typically not available at the
time of the Keck observations), a code was developed to constrain the
redshift, SN type and epoch from the available photometric data
\citep[see][for details]{2006AJ....131..960S}. In selecting candidates
for this program, the following criteria were adopted:
\begin{enumerate}
\item{}The SN photometric redshift lay at $z$$<$0.75,
\item{}The predicted phase was such that the Keck observations would
  be conducted prior to or close to maximum light.
\item{}The light curve and color were consistent with a SN~Ia,
\end{enumerate}
In practise, this last criteria is conservatively applied to minimize
the risk of selecting against SNe~Ia which may differ from the
template used in our selection code, though this reduces the fraction
of SNe~Ia observed: By the end of 2005, after 20 nights of Keck time,
58 SN candidates were observed of which 36 were confirmed as SNe~Ia
($\sim$60\%). A higher fraction of SNe~Ia could have been obtained by
a more rigorous application of the pre-selection technique; this
argues that any bias introduced by our selection is small.

\begin{figure}

\includegraphics[width=3.5in]{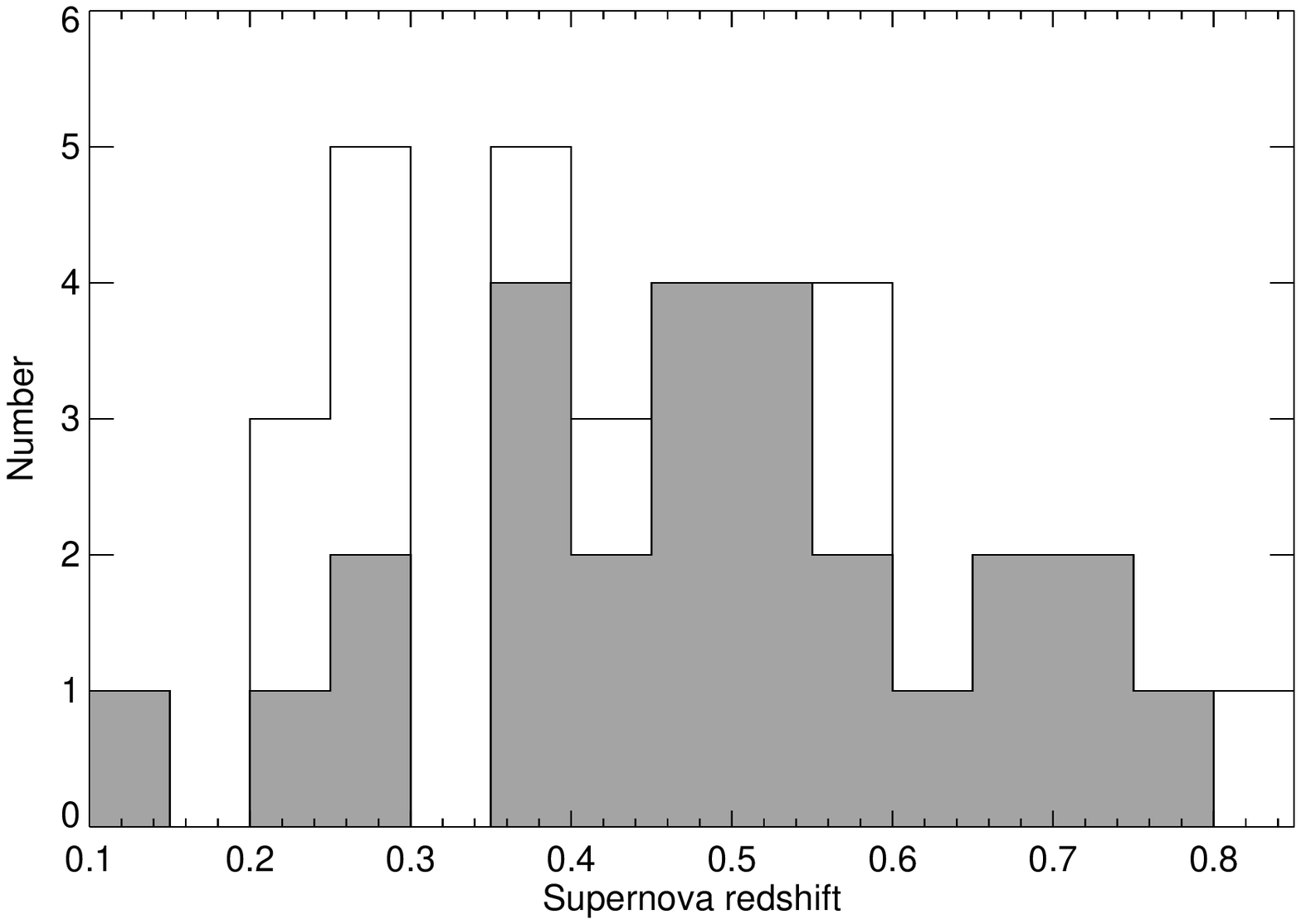}
\includegraphics[width=3.5in]{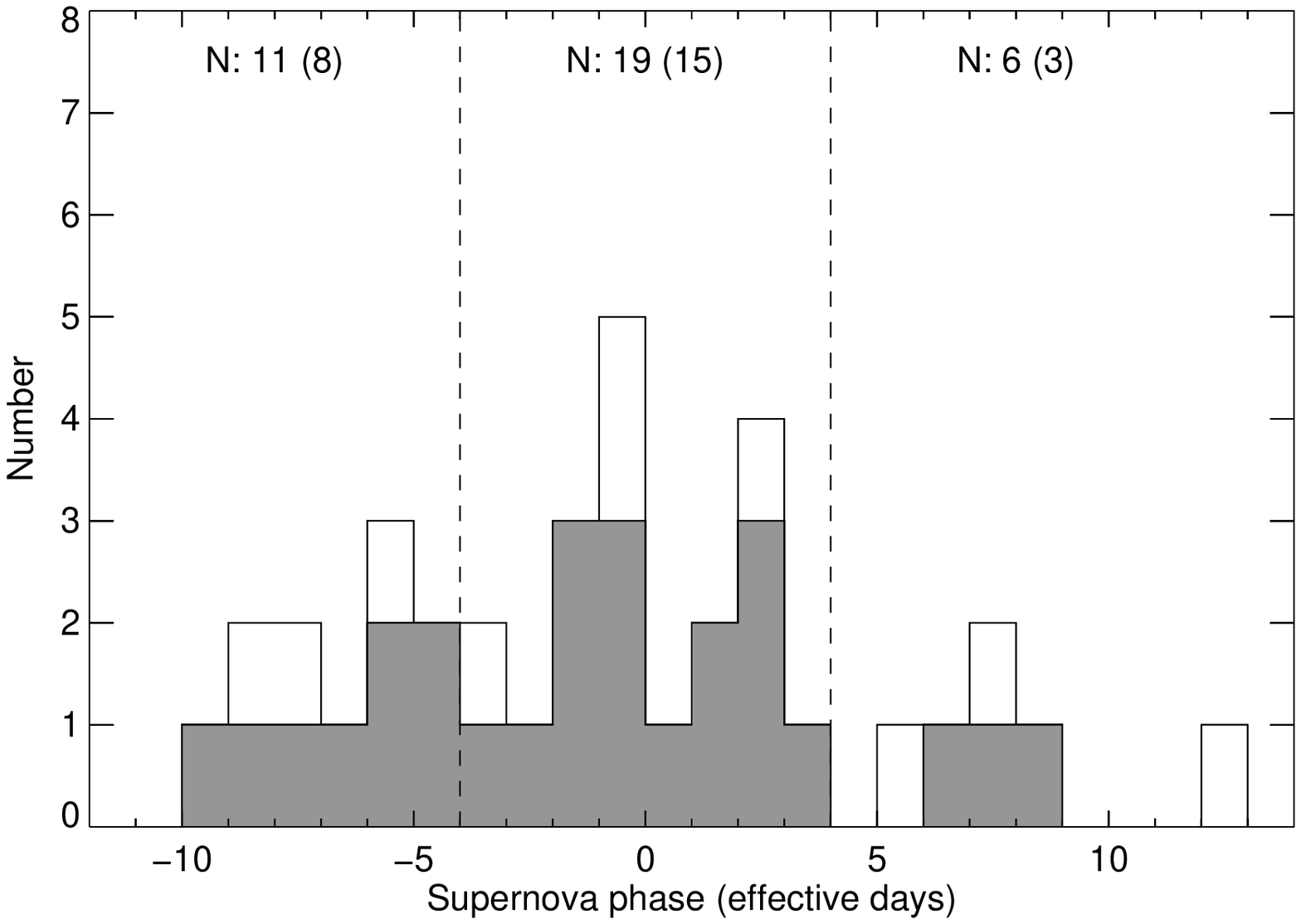}

\caption{
  Redshift (left) and phase (right) distributions of the high-redshift
  SNLS SNe~Ia studied in this paper. The vertical dashed lines in the
  phase distributions show how the SN~Ia sample is divided into ``early''
  and ``maximum-light'' spectra. Numbers (and the open histograms) refer to the primary sample of
  36 high signal-to-noise ratio spectra whereas those in parentheses (and the shaded histograms) refer to
  the sample of 26 used to construct the mean UV spectra (see text). 
  \label{fig:redshiftphasehist}}

\end{figure}

Table~\ref{tab:sninfo} summarizes the key parameters for the SNe~Ia
sampled in the Keck campaign (an independent survey of SNe~IIP was
also undertaken during this period \citep{2006ApJ...645..841N}.) We
restrict these 36 SNe to a high-quality subset, removing 3 as
spectroscopically peculiar (see $\S$~\ref{sec:mean-uv-spectrum}), 2
with a low-S/N Keck spectrum, and 5 with poor $g'$ light curve
coverage which precludes an accurate subtraction of the host galaxy
from the SN spectrum (see Table~\ref{tab:sninfo}).  These later 5 were
all observed during the SNLS ``pre-survey'' when the light curve
coverage in $g'$ was less dense, and hence our photometrically
calibrated SN spectrum will suffer a greater systematic uncertainty
(see $\S$~\ref{sec:spectr-calibr-host}). Analysis of the remaining 26
SNe~Ia forms the basis of this paper.
Fig.~\ref{fig:redshiftphasehist} displays the redshift distribution of
these high-quality SNe~Ia and the breakdown in phase at which the Keck
spectra were obtained.

For comparison purposes, we will also make use of an additional sample
of 160 spectroscopically-confirmed SNLS SNe~Ia (without high quality
Keck coverage) within the above redshift range for which good
broad-band photometric data is available. This will be used as a
control to demonstrate our SN~Ia selection is unbiased relative to the
larger SNLS population in terms of fundamental properties such as
light curve stretch, host galaxy type and location within each host.

\section{Spectroscopic Data}
\label{sec:keck-observations}

\subsection{Observations}
\label{sec:observations}

All spectroscopic observations of SNLS SNe presented here were
conducted with the 10 meter Keck-I telescope using the two-channel Low
Resolution Imaging Spectrograph \citep[LRIS;][]{1995PASP..107..375O}.
Although SNe~Ia have broad absorption features, the need to monitor
expansion velocities to reasonable precision as well as to locate
narrower diagnostic lines which would otherwise be blended in broader
features defined the need for intermediate (600 l mm$^{-1}$)
dispersion gratings. Access to the rest-frame UV over 0.2$<$z$<$0.7
was a prime goal given the sensitivity of features in this region to
metallicity, temperature and photospheric expansion velocity.

Taking these considerations into account, we adopted a spectroscopic
set up using a 400 line\,mm$^{-1}$ grating blazed at 8500\AA\ (giving
a dispersion of $\simeq1.9$\,\aperpix) on the red side and a 600 line
grism blazed at 4000\AA\ (dispersion $\simeq0.6$\,\aperpix) on the
blue side. With a 5600\AA\ dichroic, this gave a contiguous
wavelength coverage from the atmospheric limit on the blue side
through to 9400\AA\ on the red side. On some of the earliest runs in
2003, we experimented with a 300 line grism blazed at 5000\AA\
(dispersion $\simeq1.4$\,\aperpix) on the blue side and a 600 line
grating blazed at 7500\AA\ (dispersion $\simeq1.3$\,\aperpix) on the
red side with a 6800\AA\ dichroic. This provided a lower resolution
on the blue side, but a more optimal set up on the red side for the
higher redshift SNe.  However, we found the 6800\AA\ dichroic did
not permit accurate flux calibration on the blue side, because of
residual contamination in the second order of diffraction.  In very
good seeing conditions, we used a 0.7\arcsec\ slit, otherwise a
1\arcsec\ slit was used.

Observations were performed at the parallactic angle
\citep[e.g.][]{1982PASP...94..715F}. Each individual exposure was
typically between 1200-1800s, and the SN was dithered along the slit
by 3-5\arcsec\ between exposures to aid with fringe removal in the
data reduction. The telescope was refocused periodically during each
night from which an estimate of the seeing was measured.
Table~\ref{tab:observinglog} summarizes these spectroscopic
observations.

\subsection{Data Reduction}
\label{sec:data-reduction}

All spectroscopic data were processed using a pipeline developed by
one of us (MS). This pipeline uses both standard Image Reduction and
Analysis Facility (IRAF\footnote{IRAF is distributed by the National
Optical Astronomy Observatories, which are operated by the
Association of Universities for Research in Astronomy, Inc., under
cooperative agreement with the National Science Foundation.})
software as well as our own custom-written routines. LRIS is a two-arm
spectrograph and different reduction techniques are required on each
side as the red side suffers significant fringing above 7000\AA.  

Our first step is to remove the overscan level on each amplifier of
each CCD, and then subtract a master zero-frame constructed on the
afternoon of each night of observing to remove any bias pattern.  Both
sides are then divided by a normalized internal flat-field which
removes pixel-to-pixel sensitivity variations. This division also
corrects for the different gains of the LRIS CCD amplifiers. The
0.7\arcsec\ slit additionally requires division by a ``slit-flat'' or
``illumination'' flat-field as the slit illumination is quite uneven,
presumably due to milling defects in the slit. An additional
complication is that the spatial position of the slit on the CCD can
drift by several pixels over the course of the night, so the slit-flat
is shifted to best match each observation. Cosmic-rays are then
identified using LACOSMIC \citep{2001PASP..113.1420V} and removed via
interpolation from neighboring pixels.

We perform our sky subtraction following the technique of
\citet{2003PASP..115..688K}, which subtracts a two-dimensional sky
frame constructed from sub-pixel sampling of the background spectrum
and a knowledge of the wavelength distortions as determined from
two-dimensional arc comparison frames. For the blue side, we proceed
directly to the spectral extraction after sky subtraction; on the red
side we perform an additional fringe correction before extraction.

To generate the master fringe frame, we first divide every
sky-subtracted frame by the two-dimensional sky frame that was
subtracted and that caused the fringing, resulting in a
two-dimensional map of the fringe strengths per photon for each
exposure. All individual fringe maps taken in a given configuration
are averaged with sigma-clipping to form a master fringe frame,
masking out regions in each exposure that contain object flux, and
weighting each frame by the exposure time of the observation.  As the
observations are dithered, the objects are located on a different part
of the CCD in each frame, and the fringe strength at every pixel is
therefore sampled across an entire night. Finally, we boxcar smooth
the master fringe frame to increase the signal-to-noise ratio in the
fringes, which typically extend over several pixels. This smoothed,
master fringe frame is then scaled and subtracted from each red-side
exposure at the stage prior to sky-subtraction, and the
sky-subtraction then repeated on the fringe-subtracted images.

The two-dimensional frames are transformed to a constant dispersion
using comparison arc lamp images. Because the objects under study
typically have a reasonable signal-to-noise ratio, the spectral
extraction was performed by tracing the object position on the CCD and
using a variance-weighted extraction in a seeing matched aperture,
equal to 1.25 times the seeing. An error spectrum from the statistics
of the photon noise is also extracted at the same time. The wavelength
calibration of each extracted spectrum is then tweaked using the
position of the night-sky lines to account for any drift in the
wavelength solution during a given night.  We then perform an
approximate telluric and relative flux calibration using
spectrophotometric standard stars taken from
\citet{1988ApJ...328..315M}, and correcting for typical Mauna Kea
atmospheric extinction using the extinction law of
\citet{1987PASP...99..887K} and the effective airmass of each
exposure. We note that flux calibration errors introduced during this
process that vary smoothly as a function of wavelength will be
corrected by our calibration to the SN lightcurves described in the
next section.

Following optimal combination of the different exposures and relative
flux calibration of the blue and red components, our final step is to
scale the two sides into a single calibrated spectrum. We first match
the flux across the dichroic by defining narrow box filters either
side of the dichroic and scale each spectrum appropriately. We then
use a weighted mean to combine the spectra, in the process re-binning
to a constant 2\AA\ per pixel resolution. The result is a contiguous
science spectrum, together with an error spectrum representing the
statistical uncertainties in the flux in each binned pixel.

Where possible, we measure redshifts from lines in the host galaxy
spectrum. In some cases the host galaxy was too faint, or the SN too
isolated from the host galaxy, to measure a redshift from the galaxy
lines, and the redshift was determined from template fits to the SN
spectrum \citep[e.g.][]{2002AAS...201.9103H}. As a final step prior to
scientific analysis we interpolate across the positions of host galaxy
emission features (e.g., H$\alpha$, H$\beta$, \ion{O}{2}, etc.)  to
remove these from the spectra.

The SNe~Ia themselves were identified by a lack of hydrogen in their
spectra, combined with broad (several thousand km\,s$^{-1}$) P Cygni
lines of elements such as \ion{Si}{2}, \ion{S}{2}, \ion{Ca}{2},
\ion{Mg}{2}, and blends of Fe peak lines. The classification of SNe is
discussed extensively in the literature; see
\citet{1997ARA&A..35..309F} for a review, and
\citet{2005AJ....130.2788H}, \citet{2005A&A...430..843L},
\citet{2005AJ....129.2352M}, and \citet{2005ApJ...634.1190H} for
issues associated with high-redshift SN classification. SNLS as a
whole adopts the numerical classification scheme presented in
\citet{2005ApJ...634.1190H} for typing SNe: a confidence index (CI)
that a given spectrum is of a SN Ia. Three classifications denote
SNe~Ia: ``Certain SN Ia'' (CI:5; type denoted as SN~Ia), ``Highly
probable SN Ia'' (CI:4; type denoted as SN~Ia) and ``Probable SN~Ia''
(CI:3; type denoted as SN~Ia$^{\ast}$). In all of the spectra in this
paper, the relatively low-redshift compared to the SNLS sample as a
whole, coupled with the long integration times, results in high
signal-to-noise ratios ensuring the classifications are unambiguous.
All SNe~Ia presented here are CI=5 on the SNLS scheme.

\subsection{Spectrophotometric Calibration and Host Galaxy Subtraction}
\label{sec:spectr-calibr-host}

Although standard spectrophotometric stars were observed periodically
on each observing night, these provide only an approximation to a flux
calibrated spectrum.  Complexities such as differential slit-losses
(as the seeing is wavelength dependent) and uncertain atmospheric
extinction corrections require a more rigorous treatment.
Furthermore, a key issue in the interpretation of SN spectra is the
removal of underlying contaminating flux from the host galaxy, which
for some high-redshift spectra can be significant. We adopted a
procedure for removing the host galaxy spectrum that takes advantage
of the photometric $u^{\star}g'r'i'z'$ host-galaxy data available from
the SNLS, as well as the near-simultaneous SNLS $g'r'i'z'$ SN
light-curve photometry, ensuring that each spectrum is carefully
corrected to be consistent with the appropriate broad-band SN colors.

We begin by measuring the surface brightness of each galaxy (in flux
units per square arcsecond) at the position of the SN in a series of
small apertures (with radii of 1-8 pixels) for each of the
$u^{\star}g'r'i'z'$ filters. These fluxes were measured from deep
stacks with no SN light present, constructed for the purpose of
studying the properties of the SN host galaxies
\citep{2006ApJ...648..868S}.  The image quality of these stacks was
around 0.8\arcsec\ in each filter, similar to the seeing for the
majority of our LRIS spectral observations. The amount of host galaxy
light present in each SN spectrum is then estimated by interpolating
the galaxy surface brightnesses at the seeing of the observation, and
then multiplying by the product of the slitwidth ($S_W$) and the
spectral extraction aperture (1.25$\times$seeing;
$\S$~\ref{sec:data-reduction}), giving a series of $u^{\star}g'r'i'z'$
fluxes representing the host galaxy contamination in our spectra. We
then fit a series of smooth galaxy spectral energy distribution (SED)
templates to these flux data to estimate the contaminating host galaxy
spectrum.  These templates are generated by the galaxy spectral
synthesis code P\'EGASE.2
\citep{1997A&A...326..950F,1999astro.ph.12179F} rebinned to the
resolution of our SN spectra. We interpolate between neighboring
galaxy spectra to find the best-fitting template. The resulting
spectrum represents the estimate of the contaminating host galaxy
continuum in every SN spectrum.

We next use the $g'r'i'z'$ SNLS light curves and their resulting
light-curve parameterization (using the lightcurve fitter SiFTO;
Conley et al., in prep) to interpolate the SN flux at the time of each
spectral observation from nearby light-curve observations. Due to the
rolling-search nature of the SNLS and the dense light curve coverage,
most of the spectra in our main sample have $g'$ photometric data
within 2 rest-frame days of the spectroscopic observation, and all
have data within 4 days (Table~\ref{tab:sninfo}), hence the
interpolation required is small.

We next estimate the amount of SN flux that went through the LRIS slit
and was inside the spectral extraction window of 1.25$\times$seeing,
and hence is present in the reduced spectrum.  We model the SN PSF as
a two-dimensional Gaussian with $\sigma=\textrm{seeing}/2.355$ and
integrate this Gaussian through an rectangular aperture of
$S_W$$\times$(seeing$\times$1.25), where $\sigma$ is calculated at the
effective wavelengths of the $g'r'i'z'$ filters adjusting the seeing
as $\lambda^{-0.2}$.  This estimates the fraction of the total SN flux
that passed through the slit, mimicking the inclusion of differential
slit-losses. These adjusted SN fluxes are added to the host galaxy
fluxes to generate combined ``SN+host'' $g'r'i'z'$ fluxes, the
estimate of the amount of flux that passed through the slit.

We correct our combined, contiguous and flux-calibrated spectrum to
have the same colors and absolute flux level as dictated by these
``SN+host'' fluxes using a smooth interpolating multiplicative spline
function \citep[see][]{2007ApJ...663.1187H}. The size of the
correction made to the observed spectrum is usually $<$10\% and is
invariably a monotonic function of wavelength. We then subtract our
best-fitting host galaxy template, and adjust the resulting spectrum
to have the correct colors of the SN on the night of observation
(again using a spline function), giving a SN spectrum with the correct
relative and absolute flux calibration.  This final multiplicative
adjustment is small and essentially corrects the subtracted spectrum
for differential slit losses. We note that this technique would not be
possible without either the $u^{\star}g'r'i'z'$ host galaxy fluxes, or
the densely-sampled $g'r'i'z'$ SN~Ia light curves.

We estimate the uncertainty in the host subtraction process using a
Monte-Carlo simulation. We repeat the subtraction for each SN 500
times, but adjusting the different variables (seeing, host-galaxy
fluxes, SN fluxes) according to their uncertainties and assuming
normal distributions. We assume a 15\% uncertainty in the seeing, and
also simulate errors in the centering of the SN in the slit,
introduced during target acquisition or by telescope tracking errors,
of $\pm0.15$\arcsec\ (1$\sigma$). Where appropriate, we use these 500
simulated spectra to estimate errors due to host galaxy subtraction in
derived quantities (such as spectral colors) in later sections. We
also carry forward in our analysis the statistical error arising from
the photon statistics in our spectra ($\S$~\ref{sec:data-reduction}).

Our final task is to correct the SN~Ia spectrum for the effects of
extinction in the Milky-Way. To correct for Galactic extinction, we
use the dust-maps of \citet*{1998ApJ...500..525S} and a
\citet*{1989ApJ...345..245C} extinction law. We defer the more complex
discussion of extinction in the SN host galaxy and intrinsic SN color
variations until later.
  
\begin{figure}

\centerline{\includegraphics[width=3.7in]{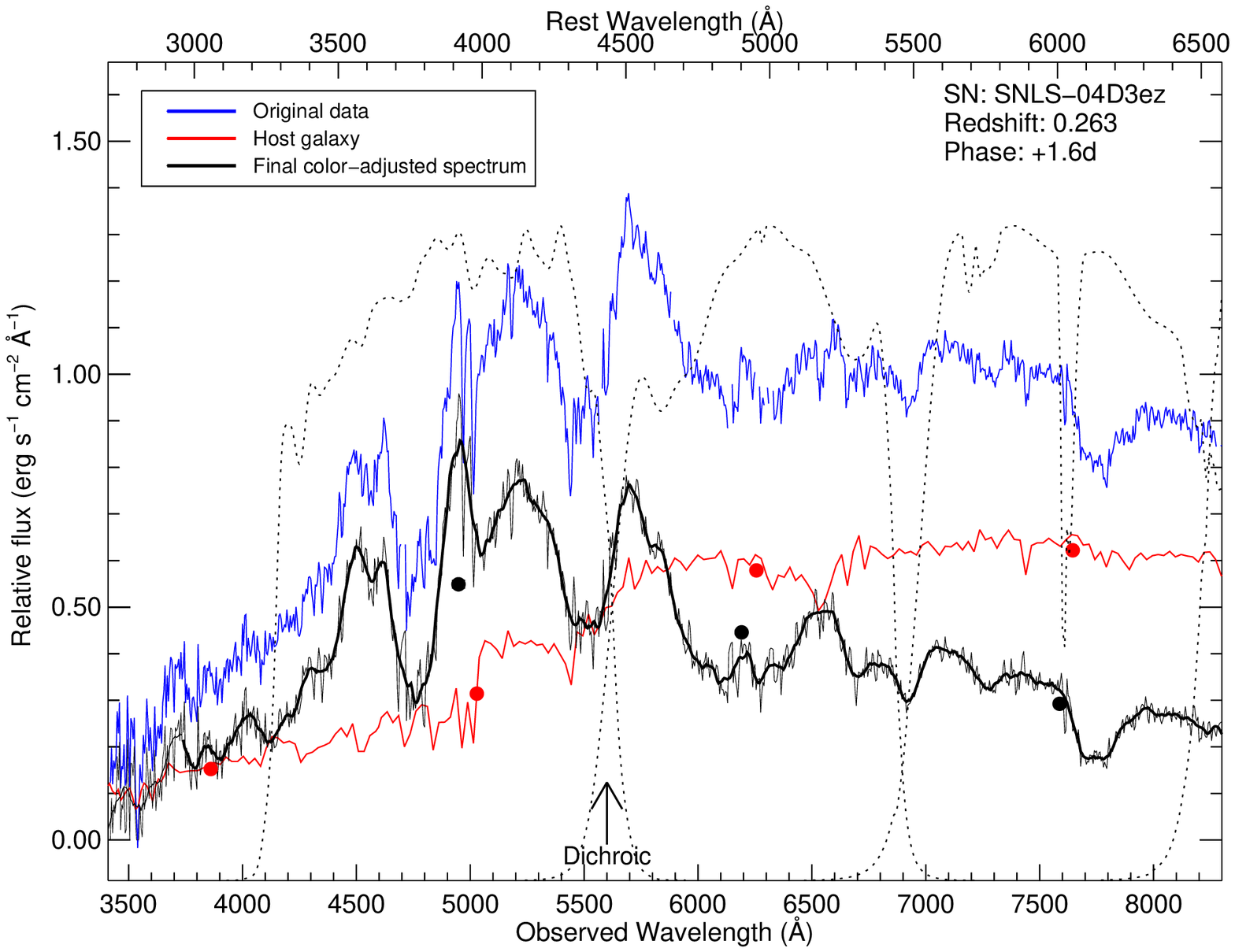} \includegraphics[width=3.7in]{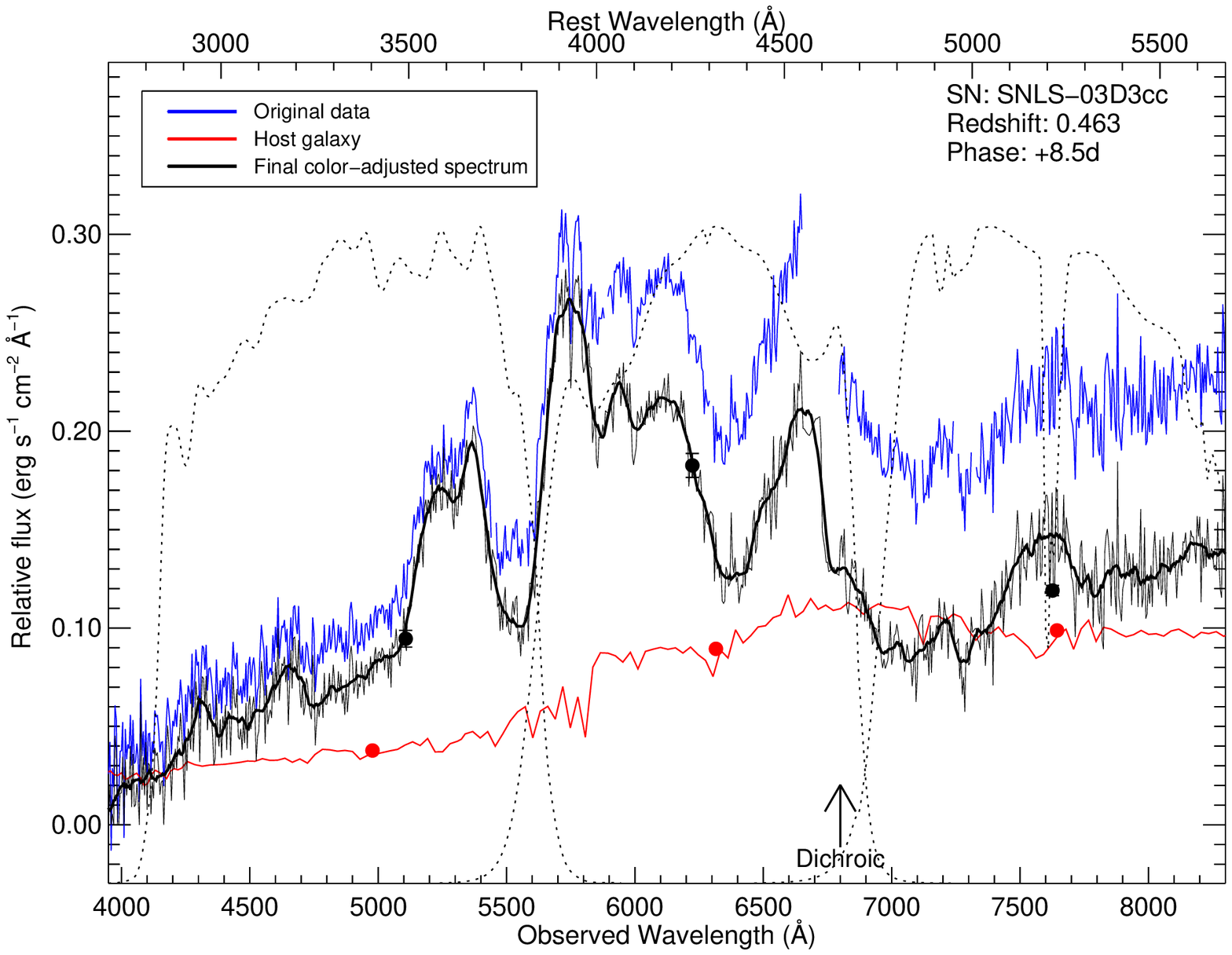}}

\caption{Examples of the host galaxy subtraction and flux calibration
  techniques for two SNe~Ia suffering host contamination. The blue
  spectrum represents the observed data, the red spectrum the
  estimated host galaxy spectrum from fits to the broad-band galaxy
  $u^{\star}g'r'i'z'$ photometry (red circles), and the black spectrum
  is the host-galaxy subtracted final spectrum color-corrected to the
  $g'r'i'z'$ SN photometry (black circles). A smoothed version of the
  final spectrum is over-plotted, the heavy line denoting the spectral
  region deemed to have the most reliable flux calibration. The
  photometric points are plotted at the effective wavelengths of the
  filters on the appropriate spectrum, and the total system response
  through the $g'r'i'z'$ filters overplotted at the observed
  wavelengths. The position of the dichroic between the two arms of
  the spectrograph is also marked. Left (a): SNLS-04D3ez at $z=0.263$.
  This SN had an increase in $i'$ of $\sim80$\% on the day of
  observation.  Right (b): SNLS-03D3cc at $z=0.463$, with an increase
  in $i'$ of $\sim200$\%. Note that due to the color correction
  applied to the final spectrum, as plotted the black spectrum differs
  slightly from the simple subtraction of the red from the blue. [See
  the electronic edition of the journal for
  Figs.~\ref{fig:eghostsub}.3--\ref{fig:eghostsub}.26]\label{fig:eghostsub}}

\end{figure}

We illustrate the results of this process for all 26 of our high quality 
SNe~Ia in Fig.~\ref{fig:eghostsub}\footnote{The online version contains 
all spectra; the printed version contains two examples.}

\section{Methodology}
\label{sec:analyses}

We now turn to the two key questions our dataset is designed to address:

\begin{enumerate}
  
\item{} Has there been significant evolution in the mean SN~Ia
  spectrum since $z\simeq$0.5 - a period of 5 Gyr corresponding to the
  past 37\% of cosmic history?

\item{} To what extent are there intrinsic spectral variations in our
  sample?  Do these give indications that SNe~Ia are a more complex
  population than assumed in cosmological studies, and might these
  variations limit the use of SNe~Ia in future cosmology experiments?

\end{enumerate}

\subsection{Theoretical Predictions}
\label{sec:theory}

We will begin to address these questions by exploring the role that
our rest-frame UV data offers as a proxy for {\it progenitor
  metallicity}. The possible effects of progenitor metallicity, either
as a function of redshift through galaxy evolution, or as a function
of environment or host-galaxy luminosity, can be explored through
blanketing and wavelength dependent features in the rest-frame UV
corresponding to $\lambda\lambda$ 2900-3500\AA.  At z$>$0.2, this
portion of the rest-frame spectrum is well-sampled in our LRIS data.
In making the assumption that our UV data are indicative of
metallicity effects, it should be realized we are relying largely on
theoretical studies; the UV spectral region is poorly explored
observationally.  First we review the various theoretical
expectations.

\citet*{1998ApJ...495..617H} argue that direct traces of the progenitor
metallicity can best be seen in the unburned SN layers which are only
observable significantly before maximum light.  However, they also
predict that an increase in progenitor metallicity will cause an
increase in the amount of $^{54}$Fe synthesized in the explosion, and
this will result in an increase in line opacity in the UV region which
may be observable at maximum light.  The degree of mixing in the
explosion complicates the interpretation, however.  The net effect
predicted by H\"oflich et al. is that an increased metallicity will
result in an {\it increase} in the UV pseudo-continuum at maximum
light.

\citet{2000ApJ...530..966L} start with the findings of
\citet{1998ApJ...495..617H} and examine the spectroscopic implications
in greater detail.  Using the starting isotopic distribution of the W7
model (Nomoto et al. 1984), they simultaneously change the progenitor
metallicity in the unburned C+O region and increase the amount of
$^{54}$Fe in the partially burned region.  They find two effects -- a
shift in the wavelength of UV features redward with decreasing
metallicity, and a simultaneous increase in the level of the UV
pseudo-continuum.  They argue that as metallicity decreases so the line
opacity decreases with the result that lines form deeper in the
atmosphere, and therefore from a lower velocity region. A lower
metallicity also decreases the cooling producing a higher temperature
and bluer colors.  We note that such a color difference is {\it
  opposite} to the effect predicted by H\"oflich et al.  However,
Lentz et al. caution that the overall UV flux level is not necessarily
a good indicator of metallicity, as it is dependent on many variables
such as the temperature, density and velocity of the C+O layer.

\citet*{2003ApJ...590L..83T} argue for much more dramatic changes in
SN~Ia physics with metallicity.  They argue that synthesized $^{56}$Ni
mass should be linearly proportional to progenitor metallicity.  Since
the decay of $^{56}$Ni drives the luminosity, SNe~Ia in high
metallicity environments should be less luminous.  This is because
stars from higher metallicity environments will end up with larger
mass fractions of $^{22}$Ne and $^{56}$Fe after helium burning.  Since
these isotopes have excess neutrons, the authors argue that in these
cases fewer radioactive elements are produced during the process of
burning to nuclear statistical equilibrium during a SN~Ia.  Note that
these results are in sharp contrast to other studies which found no
significant increase in $^{56}$Ni with increasing metallicity
\citep[e.g.][]{1998ApJ...495..617H,1999ApJS..125..439I}

Finally, we note that various authors predict that the SN~Ia rate
should be affected by metallicity, although they do not make explicit
predictions about the resulting effects on SN~Ia properties.
\citet{1998ApJ...503L.155K} argue that in very low metallicity
environments ($[Fe/H]< -1$) the white dwarf wind that they believe is
essential for producing SNe Ia will be inhibited, thus leading to
fewer SNe~Ia.  \citet{2000A&A...362.1046L} find that metallicity
differences should alter the range of progenitor masses that produce
SNe~Ia.

In summary therefore, theory cannot yet offer us a clear consensus as
to the effects of metallicity on the UV properties of SNe~Ia. Indeed,
there is disagreement not only about which effects are the most
important, but also about the sign of any possible effect.  This is a
very challenging theoretical problem hindered by correlations between
the wanted effect of metallicity and other correlations such as the
viability of certain progenitor systems, the explosion mechanism and
radiative transfer in an atmosphere under a variety of mixing
conditions. Full simulations of all these effects may soon become
feasible, but substantial campaigns will still be needed to track
statistical shifts with metallicity. We therefore undertake an
empirically-motivated study of the properties of supernovae in the UV,
and note where the observations agree or disagree with certain
theoretical studies. In this respect, it is helpful to have some
theoretical basis for making the empirical measures.

\begin{figure}

\centerline{\includegraphics[width=4in]{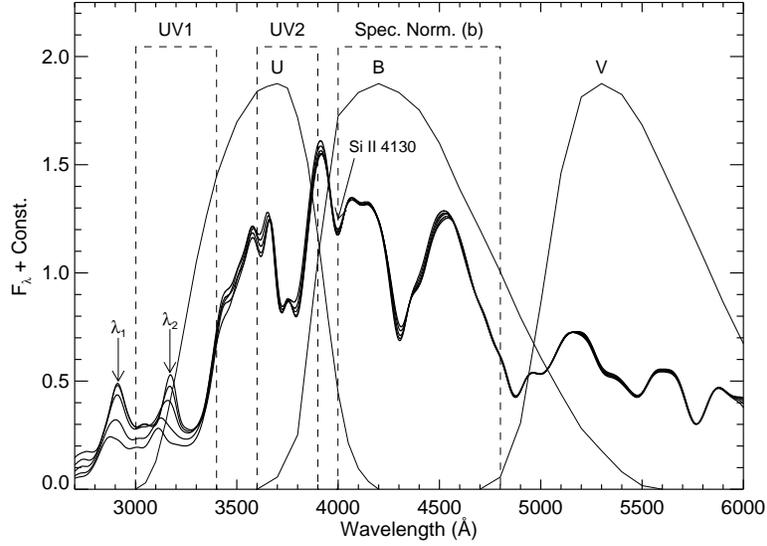}}
\caption{Maximum light SN~Ia spectra from \citet{2000ApJ...530..966L}
  at a variety of different metallicities (top to bottom: 0.1, 0.3,
  1.0, 3.0 and 10.0 times solar metallicity). Over-plotted are the
  standard \citet{1990PASP..102.1181B} $U$ and $B$ filter responses,
  as well as the UV box filters (``UV1'' and ``UV2'') and
  spectroscopic normalizing filter (``b'') used in the analysis.
  Features thought to represent possible metallicity diagnostics
  ($\lambda_1$, $\lambda_2$) are also marked, as well as the position
  of the blue-shifted \ion{Si}{2} 4130\AA\ absorption
  feature.\label{fig:col_lentz}}
\end{figure}

Fig.~\ref{fig:col_lentz} illustrates, for the range of metallicities
sampled by \citet{2000ApJ...530..966L}, that the bulk of the spectral
changes would clearly be detectable within our rest-frame wavelength
range. The blanketing and wavelength shifts apparent in
Fig.~\ref{fig:col_lentz} suggest the use of both photometric and
spectroscopic diagnostics. As high quality UV spectra are in scarce
supply at low redshift, we can best make meaningful comparisons using
rest-frame $U-B$ measurements. As our spectra do not always extend to
the full redward end of the standard $B$ filter, we will also adopt a
top-hat filter extending from 4000 to 4800\AA\ which we will refer to
as $b$.  Pseudo-photometric ultraviolet measures can be constructed
from our spectral data using two top-hat filters, UV1 and UV2,
centered around regions where the strongest trends are predicted.
Although these various photometric measures degrade the information
content of our spectra, they have the considerable benefit of being
consistent over a wide range in redshift without making any
assumptions about the k-correction. The metallicity-induced spectral
shifts discussed above can likewise be tracked by measuring the rest
wavelengths of two UV features at $\lambda\lambda$2920 and 3180 \AA\ .
We will refer to these indicators as $\lambda_1$ and $\lambda_2$.
Both the photometric and wavelength features introduced above are
marked in Fig.~\ref{fig:col_lentz}.

\subsection{Suitability of the Keck SNe~Ia Sub-sample for Detailed Studies}
\label{sec:suitability-keck-sne}

If our conclusions concerning the properties of $z\simeq$0.5 SNe~Ia
are to be relevant for the role of SNe~Ia as cosmological probes, our
sample of SNe~Ia should be a representative subset of those being used
for probing dark energy. Accordingly, first we will verify whether
this is the case by comparing the distribution of properties with
those for the parent SNLS SN~Ia sample. This parent sample (N=160)
includes those presented by \citet{2006A&A...447...31A}, as well as
further events up to May 2006, restricted to lie within our chosen
redshift range 0.15$<z<$0.7. This is the redshift range over which
SNLS suffers least from Malmquist-type effects in the properties of
SNe chosen for spectroscopic follow-up, and the broad SNLS population
is reasonably unbiased over this redshift range
\citep[e.g.][]{2006A&A...447...31A,2006AJ....132.1126N}.

\begin{figure}

\centerline{\includegraphics[width=4in]{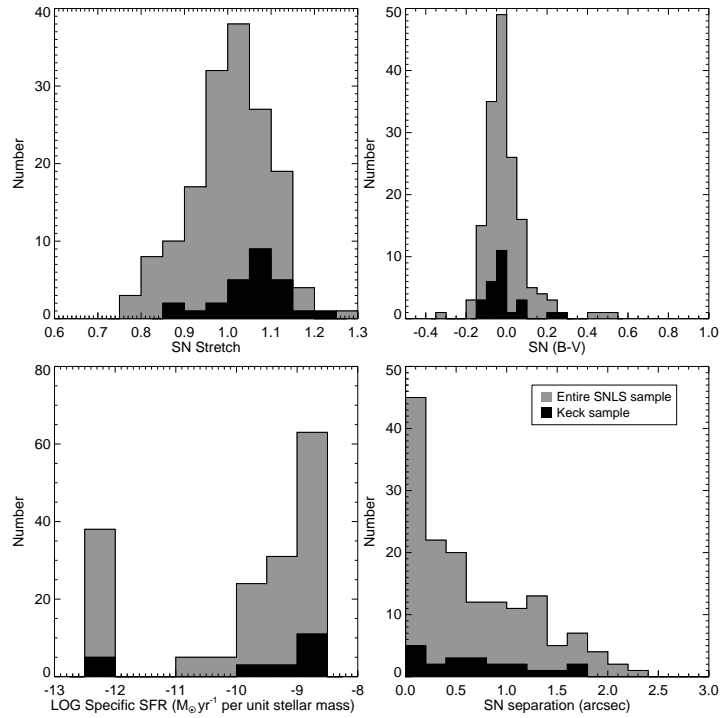}}
\caption{Comparison of SN stretch, SN B-V color, host galaxy specific
  star formation rate and SN/galaxy projected separation for the Keck
  SN~Ia sample (black histogram; present paper) and the entire SNLS
  sample within the appropriate redshift range discovered over the
  same period (grey histogram). Galaxies with undetected star
  formation were placed at 10$^{-12}\, M_{\odot}$ yr$^{-1}$ per unit
  stellar mass.\label{fig:samplecomparison}}
\end{figure}

Fig.~\ref{fig:samplecomparison} compares the distribution of SN~Ia
stretch, location and host galaxy properties for the two samples.
There is no evidence that our selection criteria ($\S$2) have
significantly biased the SN~Ia population chosen for intensive study
relative to the larger SNLS population.  Kolmogorov-Smirnov (K-S)
tests indicate the Keck and SNLS sample distributions in all four
parameters are consistent with being drawn from the same distribution;
the exception is the stretch distribution where the K-S test indicates
the distributions are possibly different due to a marginal deficiency
of low stretch events in the Keck sample.  This can be easily
understood as these events are both fainter (making the Keck
spectroscopic observations more demanding) and have faster light
curves (making a pre-maximum or at maximum light observation less
likely, a criterion when selecting SNe for Keck followup;
$\S$~\ref{sec:selection-cfhtls-sne}) than higher-stretch events.
Otherwise, the larger scale environs in which our SNe~Ia occurred seem
quite representative of the parent population, for example in the
specific star formation rate, the separation from the host galaxy, and
(not shown in Fig.~\ref{fig:samplecomparison}) the absolute magnitude
of the host galaxy.

\subsection{Local UV Spectra}
\label{sec:local-data}

In order to examine possible evolutionary effects we require a local
baseline UV spectrum. For this, we turn to the analysis of
\citet*{2002PASP..114..803N} which provides a phase-dependent template
SN~Ia spectrum which has become the standard reference used
extensively, for example, in the calculation of cross-color
$k$-corrections in deriving SNe~Ia luminosities at high redshift.
Further details of its construction can be found in the original
article. Note that though improved spectral templates now exist
\citep[e.g.][]{2007A&A...466...11G,2007ApJ...663.1187H}, these
typically exploit high-$z$ spectra in their construction, whereas
Nugent et al. use solely local SN~Ia data.

The UV portion of this mean local spectrum is based on very few local
events studied by either the IUE or HST (see Table~1 in
\citet{2002PASP..114..803N}), and significant uncertainties remain at
the shortest wavelengths. Unfortunately, unless STIS is resuscitated
on the upcoming HST Servicing Mission, there is no immediate prospect
of improving this situation. A particular concern is the fact that
within a week of peak brightness only three SN~Ia (comprising of a
total of five spectra) contribute to the UV template. They are
SNe~1981B, 1990N and 1992A. These SNe have stretches of 0.929, 1.074
and 0.819 measured using SiFTO, i.e. within the range sampled here
(Fig.~\ref{fig:samplecomparison}). However, their hosts, classified as
SAB(rs)bc (NGC~4536), SAB(rs)bc (NGC~4639) and SA0 (NGC~1380)
respectively, are possibly biased towards large early type systems.

\section{Results}
\label{sec:results}

\subsection{Tests for Progenitor Evolution: the Mean UV SN~Ia Spectrum}
\label{sec:mean-uv-spectrum}

We begin by comparing the mean spectrum of the intermediate-redshift
Keck SN~Ia sample with the local \citet{2002PASP..114..803N} template.
Earlier work comparing the properties of low and high-redshift SN~Ia
spectra
\citep[e.g.][]{2000ApJ...544L.111C,2005AJ....130.2788H,2006AJ....131.1648B,2007A&A...470..411G,2007bronder}
has concentrated largely on comparisons of optical spectral features
involving lower-S/N spectra. In general, these studies find that
equivalent widths and ejection velocities of optical spectroscopic
features measured on high-redshift SNe~Ia have similar distributions
to those observed locally. No study has yet found evidence for
evolution in the optical spectroscopic properties of SNe~Ia with
redshift.

In this paper, we concentrate on the properties of the mean UV
spectra. As discussed earlier, models suggest metallicity variations will
have greatest effect in this wavelength region and thus our sensitivity
to evolutionary effects should be maximized.  We separate the distant 
SNLS events into three groups
(Fig.~\ref{fig:redshiftphasehist}): those observed prior to maximum
light ($t<$-4 rest-frame days), those observed close to maximum light
(-4$<t<$+4 days), and those observed after maximum light ($t>$+4
days).  We exclude 3 spectra, labeled (f) in Table~\ref{tab:sninfo},
which we classify as peculiar. One is the super-Chandrasekhar event
SNLS-03D3bb \citep{2006Natur.443..308H} and the other two (SNLS-05D1hk
and SNLS-03D4cj) have peculiar spectra similar to that of SN1991T-type
\citep{1992ApJ...384L..15F}.  (A further 7 events, labeled (d-e) in
Table~\ref{tab:sninfo}, were discarded as described in
$\S$~\ref{sec:selection-cfhtls-sne}.) At this stage, our comparisons
will involve spectra uncorrected for host galaxy extinction and/or
intrinsic color variations.

\begin{figure}

\plotone{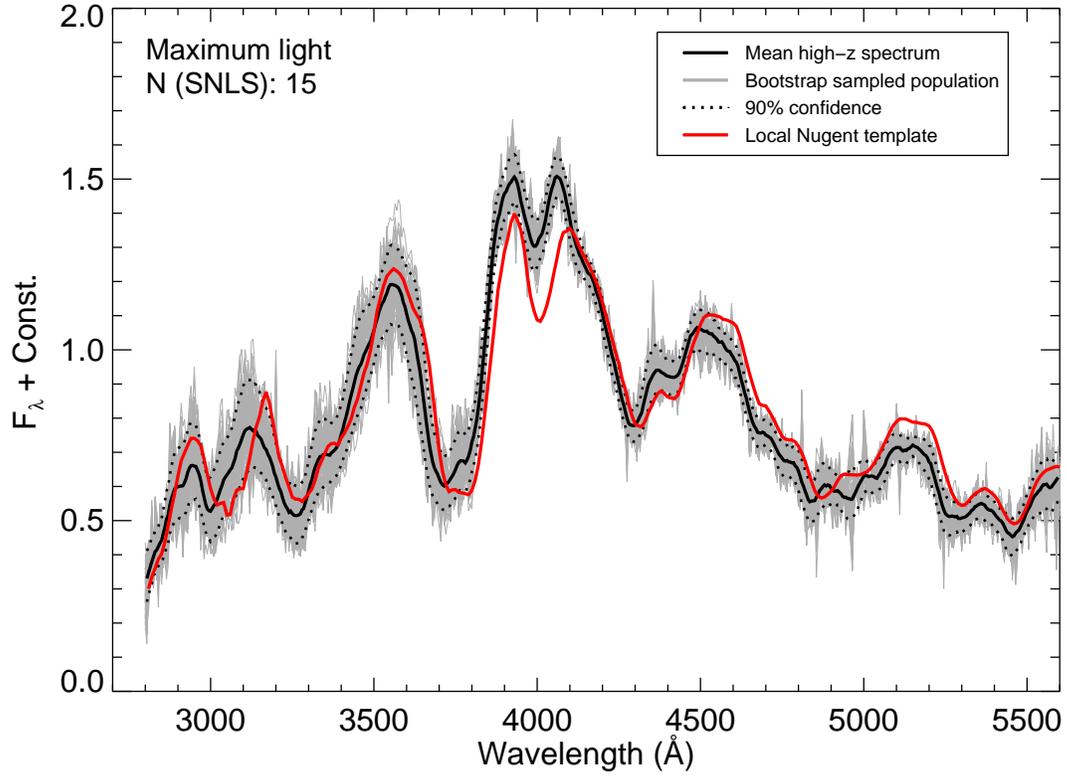}
\caption{
  The mean high-redshift maximum light (effective day $< \pm$ 4 days) 
  rest-frame UV SN~Ia spectrum compared to the local average template of
  \citet{2002PASP..114..803N}. Over-plotted in light grey are 100
  bootstrap-resampled mean spectra drawn from the high-redshift
  population; the dotted lines show the region containing 90\% of
  this distribution. The local template has been color-adjusted to
  match the high redshift data.
  \label{fig:mean_max}}

\end{figure}

\begin{figure}

\plotone{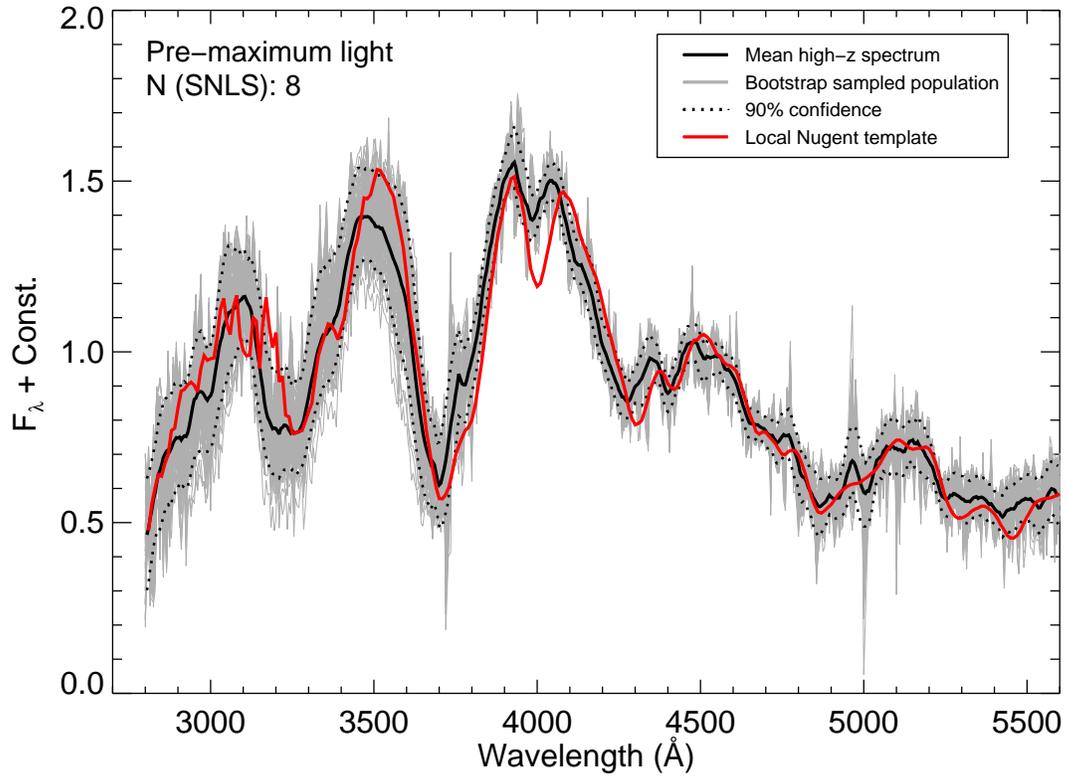}
\caption{ As Fig.~\ref{fig:mean_max}, but showing the pre-maximum
   (effective day $<$-4 days) high-redshift spectrum versus the local template.
  \label{fig:mean_early}}

\end{figure}

Our 26 high-quality SNe~Ia consist of 8 pre-max, 15 maximum light, and
3 post-max spectra. In creating a composite spectrum, flux-calibrated
spectra are normalized using the top-hat $b$ filter introduced earlier
which runs from $\lambda\lambda$4000-4800\AA\ (see
Fig.~\ref{fig:col_lentz}). We use a clipped mean and determine the
error on the mean by bootstrap resampling.  To facilitate a meaningful
comparison we color-adjust the local spectrum so that its $U-B$ color
matches that of the mean high redshift sample.
Figs.~\ref{fig:mean_max} and \ref{fig:mean_early} show the
maximum-light and pre-maximum high-$z$ mean spectra compared against
the local template at similar phases. For convenience, we present in
Tables~\ref{tab:mean_spectra_early} and \ref{tab:mean_spectra_max},
the early and maximum light spectral energy distributions with upper
and lower 90\% confidence limits. For the local template, due to the
very small number of SNe~Ia with rest-frame UV spectra
($\S$~\ref{sec:local-data}), a similar bootstrap resampling technique
to asses the uncertainty in the local template is not meaningful.

Examining Fig.~\ref{fig:mean_max}, it is clear that the
spectrum at $\overline{z}\simeq$0.5 is reasonably similar to its local
equivalent.  The agreement, particularly longward of 4000\AA, is
reassuring, indicating that there has been no significant evolution in
the mean SN~Ia spectroscopic properties over the past 5~Gyr. However,
some systematic departures are seen, for example, in the
\ion{Si}{2}/\ion{Co}{2} doublets at $\simeq$4150\AA\ and in the
positions of the UV features around 2900-3200\AA.

Do these UV changes represent systematic differences in the progenitor
properties between $z\simeq$0.5 and 0? In this respect it is helpful
to consider the error in the $z\simeq$0.5 mean spectrum (shaded in
Fig.~\ref{fig:mean_max}).  Here we see that the local
spectrum lies within the 90\% confidence boundary in the distribution
of $z\simeq$0.5 spectra. Thus it is quite possible that the UV
differences seen between high and low redshift arise as a result of
statistical variations occurring within a non-evolving population,
especially considering the number of local SNe~Ia with rest-frame UV
coverage is very small.

Turning to the pre-maximum light spectra
(Fig.~\ref{fig:mean_early}), both \citet{1998ApJ...495..617H} and
\citet{2000ApJ...530..966L} argue that differences in the UV spectrum
due to changes in the progenitor metallicity should be larger at
earlier times, where the unburned outer layers of the white dwarf play
a larger role in shaping the appearance of the spectrum.  Although the
agreement does indeed seem to be poorer, the local template is
considerably uncertain at this phase; we conclude that the most
likely explanation for the differences we see is that the local
template is unrepresentative in the far UV.

However, the differences that we do see in the UV data, whether due to
evolution or simply as a result of intrinsic scatter in the
population, are much larger than the effects predicted by
\citet{2000ApJ...530..966L} for quite substantial changes in
progenitor metallicity.  It should be noted that in these models,
luminosity and kinetic energy were fixed, whereas our observations
presumably span a considerable range in both of these variables.
Lentz et al. also predict systematic wavelength shifts of the
diagnostic features identified as $\lambda_1$ and $\lambda_2$ in
Fig.~\ref{fig:col_lentz}, the study of which should be independent of
reddening. We will defer consideration of these spectroscopic features
until $\S$~\ref{sec:spectroscopic_features}.

At longer wavelengths, the local data is more reliable.  Here, several
features of the pre-maximum spectrum deserve comment
(Fig.~\ref{fig:mean_early}).  Compared to the low-z template, the
high-z average spectrum has a shallower \ion{S}{2} 5400\AA\ feature,
and a shallower \ion{Si}{2} 4130\AA\ feature.  In addition the
3700\AA\ Ca+Si feature is narrower and shows a greater degree of
splitting into the separate Ca and Si components in the high-z
composite spectrum.  More high-z SNe~Ia belong to what
\citet{2006PASP..118..560B} call the ``shallow Si'' group --- SNe such
as SN~1991T, which have a narrower zone of intermediate mass elements.
These results agree with the findings of \citet{2001ApJ...546..734L}
who note that, while a significant fraction of SNe~Ia show such
behavior at early times, they become more normal at maximum light.
This supports the claims of \citet{2006ApJ...648..868S}: at high
redshift, where the fraction of star forming hosts is higher, more
``prompt'' SNe are found and these tend to have broader light curves.

\begin{figure}

\plotone{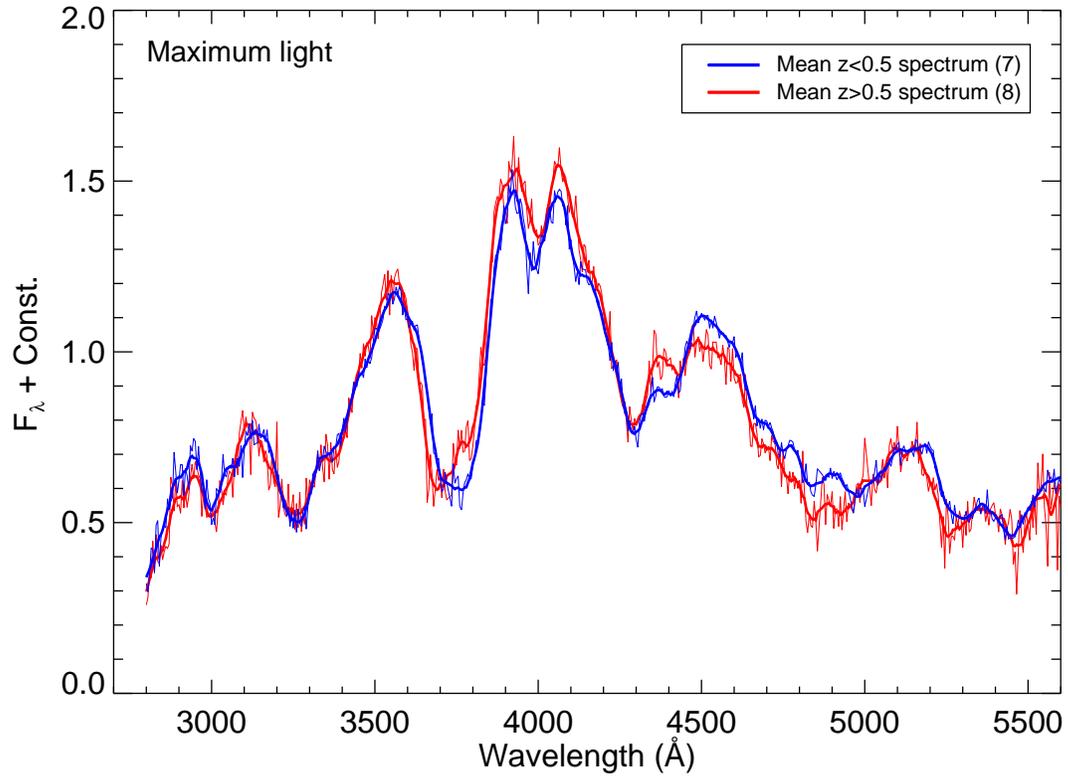}
\caption{The mean high-redshift maximum light rest-frame UV SN
  spectrum split into two redshift intervals below (black) and above
  (red) $z$=0.5.\label{fig:mean_max_z}}
\end{figure}

The lack of a comprehensive local sample with comparable quality to
the Keck data extending into the UV clearly limits any precise tests
for spectral evolution. This impasse raises the question of whether it
might be more profitable to search for evolutionary trends within the
Keck sample itself. Fig.~\ref{fig:mean_max_z} shows the mean maximum
light spectra compiled for two Keck sub-samples, split into two
redshift ranges at $z$=0.50. The mean redshifts of the two subsamples
are $z$=0.36 (N=7) and $z$=0.58 (N=8). The agreement is striking, with
only marginal evidence for differences in blanketing at the shortest
wavelengths.  Further progress in this respect may be possible with
the ongoing SDSS SN~Ia survey (which samples $\overline{z}\simeq$0.2)
and the higher-redshift SNLS SN~Ia spectra at $z\simeq0.6-1.0$.

In summary, no convincing evidence is available for a {\it systematic}
change with redshift in the UV spectra of SNe~Ia. The bulk of the
differences seen probably arise from a natural dispersion within the
sample and it seems reasonable to assume this dispersion is present at
all epochs. In the next section we explore the possible origin of this
UV dispersion.

\subsection{Reality of the UV Dispersion}
\label{sec:disp-uv-reality}

In this section we first consider the physical reality of the
dispersion seen in the UV spectra at maximum light. The
bootstrap-resampled variations (Figs.~\ref{fig:mean_max} and
~\ref{fig:mean_early}) indicate a greater degree of scatter at
$\lambda<3700$\AA\ than in the optical around $\lambda\sim5000$\AA.
However, both the effects of dust extinction and the size of the
calibration uncertainties are unaccounted for in these comparisons.
Could the increased dispersion simply arise as a result of one or
both of these effects and not represent a true physical variation
amongst our SNe?

We first discuss the effects of dust extinction.

\subsubsection{Effects of Dust Extinction}
\label{sec:dust}

An explanation for the UV variations is the presence of varying amount
of interstellar dust, either in each host galaxy or along the line of
sight to the observer.  Correcting the properties of SNe~Ia for
extinction is currently one of the most pernicious problems in their
use as cosmological probes \citep[e.g., see discussion
in][]{2007arXiv0705.0367C}. Earlier analyses
\citep{1996ApJ...473...88R,1998AJ....116.1009R, 1999ApJ...517..565P,
  2000ApJ...536...62R, 2003MNRAS.340.1057S, 2003ApJ...598..102K},
based on the first generation of SN~Ia surveys, examined the range of
likely extinction using both SN colors and host-galaxy morphologies as
markers.  Generally only modest levels of extinction ($A_V<0^m.2$)
were seen at high-redshift across the full range of host types, with
no evidence of any systematic change with redshift in the samples used
for cosmological analyses.

In the case of the present SNLS sample, the key diagnostic is the
rest-frame $B-V$ color estimator ``$c$''
\citep{2005A&A...443..781G,2007A&A...466...11G}, essentially the $B-V$
color measured at maximum light.  We wish to test whether dust is the
primary cause of the UV variations seen in the Keck spectra. This can
be done by considering, statistically, correlations between the UV
spectra and $c$. Reddening measures for each SN are not directly
output by the SNLS analysis since $c$ is treated as an empirical
variable when performing the cosmological analysis
\citep{2006A&A...447...31A}; no physical model is assumed for the
color variations.  However, we can readily test the hypothesis that
the bulk of the UV dispersion arises from reddening that is linked to
the $B-V$ variations.

We investigate two methods for color-correcting our SNe~Ia spectra.
The first is to use a standard reddening law
(\citet{1989ApJ...345..245C}, CCM) and extinction-correct each
individual Keck spectrum using that SN's $c$ measurement and a value
of $R_B=4.1$, suitable for standard Milky-Way type dust.  The results
from this process are not particularly sensitive to either the exact
$B-V$ color used as the fiducial/zeropoint SN~Ia color, nor to the
exact value of $R_B$ used. Our second approach is similar but uses the
SALT \citep{2005A&A...443..781G} color law together with the $c$
estimated from the light curve fits. As this color law was trained on
local SN spectra and colors, it includes not only the effect of dust
extinction but may also partially account for intrinsic variations
between SN optical colors and UV fluxes unrelated to dust.  The CCM
and SALT laws have a similar form in the optical, but differ in the
near-UV, where the SALT law implies more ``extinction'' in the UV than
the CCM law \citep{2007A&A...466...11G}.  To test the validity of
these approaches we split those Keck spectra studied at maximum light
into two sub-samples according to whether the rest-frame maximum light
optical color is redder or bluer than a $c$=-0.057.

\begin{figure}

\centerline{\includegraphics[width=4.25in]{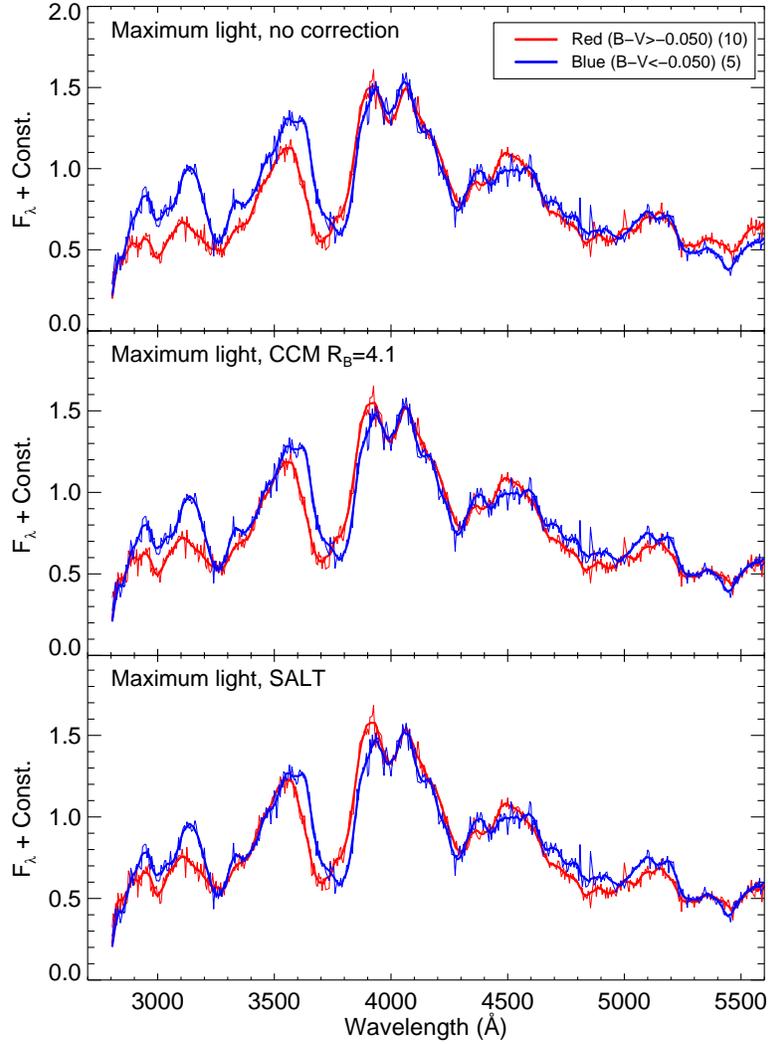}}

\caption{A test of the validity of applying a color correction to
  those Keck spectra sampled at maximum light. The top panel shows the
  mean {\it observed} spectrum for two subsamples split according to
  the rest-frame $B-V$ color at maximum light; red corresponds to SNe
  with $B-V>-0.057$, blue to those with $B-V<$-0.057. The middle panel
  shows the same comparison after applying a
  \citet{1989ApJ...345..245C} reddening correction to each individual
  spectrum using an $E(B-V)$ estimated from the light curve fits, and
  $R_B=4.1$. Varying $R_B$ has a negligible effect on this comparison.
  The lower panel shows the comparison after correcting individual
  spectra using the SALT color law
  \citep{2005A&A...443..781G}.\label{fig:mean_max_col}}
\end{figure}

Examining the mean UV spectra for the two subsets
(Fig.~\ref{fig:mean_max_col}), we find that, before any correction,
optically bluer SNe~Ia do indeed have more flux in the UV as expected
if they are less extincted. When the spectra are individually
color-corrected using a CCM law and then combined by subset, the mean
UV spectra agree better though the entire color difference is not
corrected shortward of 3500\AA. The approach using the SALT color law
performs better in this UV region, and provides a good agreement
across our entire wavelength range, but is unable to correct all the
differences seen. Given the slightly superior performance of the SALT
color law, we adopt this approach for color-correcting our spectra in
the remainder of this paper.

\subsubsection{Effects of Calibration Uncertainties}
\label{sec:effect-calibr-uncert}

Our host galaxy subtraction and spectral calibration procedures are
described in $\S$~\ref{sec:spectr-calibr-host}. For each
host-subtracted and photometrically-calibrated SN spectrum we also
record a series of 500 Monte-Carlo simulated spectra generated by
randomly varying the SN fluxes, host galaxy fluxes, seeing and SN
position in the slit according to the observational errors. 

For each of the SN studied, we estimate the 1-$\sigma$ uncertainty as
follows. In each wavelength bin for a given SN, we order the 500
random spectra and find the range in flux that encompasses 68\% of the
population and assign this flux range as the 1-$\sigma$ error in this
wavelength bin. This is repeated in every wavelength bin for all the
spectra, providing the error from the observational uncertainties for
each SN spectrum as a function of wavelength.

\begin{figure}
\plotone{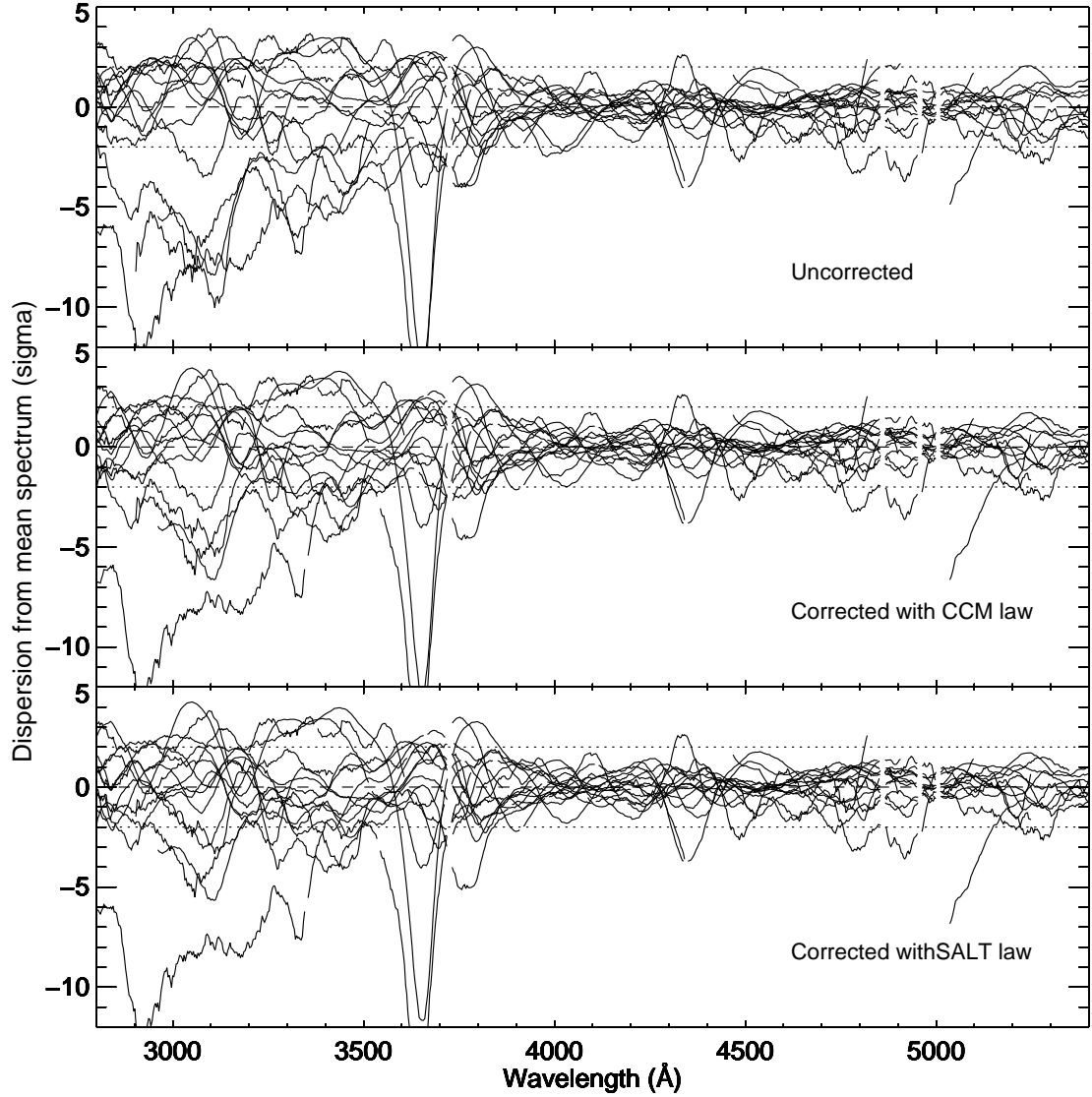}
\caption{The deviation of the 15 maximum-light spectra from the mean
  spectrum in $\sigma$-units derived from their error spectra
  $\S$~\ref{sec:effect-calibr-uncert}. The top panel shows the
  uncorrected spectra, the middle panel the spectra corrected with the
  CCM dust law, and the lower panel spectra corrected with the SALT
  color law. The horizontal lines denote $\pm$2-$\sigma$
  deviations.\label{fig:meandisp}}
\end{figure}

We can then assess whether the scatter that we see about the mean
spectrum is significant. For every spectrum, we calculate the residual
from the mean spectrum and divide by the appropriate error spectrum.
The result is the deviation from the mean for each spectrum in units
of sigma. Smoothed versions of these are plotted in
Fig.~\ref{fig:meandisp}; the top panel shows the deviation for the
uncorrected spectra, the middle and lower panels show the deviations
for spectra corrected using the CCM and SALT laws as described in
$\S$~\ref{sec:dust}.

In general, the scatter is markedly larger in the UV
($\lambda<4000$\AA) than in the optical region. In the optical, the
spectra appear well-behaved, with most of the deviations $<$2$\sigma$,
consistent with a mean spectrum showing little intrinsic dispersion
over the broad wavelength range.

These tests and those of the previous section demonstrate that most of
the difference in the \textit{average} continuum levels seen between
the mean UV spectra of optically blue and red SNe~Ia can be corrected
using existing techniques such as the CCM dust extinction law or the
SALT SN~Ia color law. However, these techniques can only reduce, and
not eliminate, the scatter observed, which is significantly larger
than that expected from a consideration of the observational
uncertainties. We thus conclude that the bulk of the dispersion
is intrinsic to our SNe and investigate its origin in the next sections. 

\subsection{Physical Correlations within the UV Dispersion}
\label{sec:uv-disp-behav} 

\subsubsection{Photometric Comparisons}
\label{sec:photometric-comparisons}

Given we have established that the UV variations are larger than those
expected from photometric errors and are still present after attempts
to correct for color differences, it is important to consider the
extent to which the intrinsic dispersion we see is consistent or
otherwise with that observed locally. Although no comparable local
spectroscopic dataset exists, \citet{2006AJ....131..527J} have
analyzed a homogeneous photometric database of 44 local SNe~Ia and
examined the $U-B$ color dispersion at maximum light in terms of both
the light curve stretch factor $s$ and redder colors such as $B-V$.
As in the higher redshift data, there is still the complication of
separating the effects of host galaxy dust extinction and intrinsic
color effects. In the Jha et al. study, host galaxy extinction
estimates were available for a subset of the data yielding intrinsic
$U-B$ and $B-V$ colors. The increased intrinsic dispersion seen in the
$U-B$ vs. stretch relationship with respect to that in $B-V$ can thus
be attributed primarily to an increased dispersion in the $U$ band of
$\sigma_U\simeq$0.12 ($\simeq$12\%).

\citet{2006A&A...447...31A} also address the question of the intrinsic
dispersion in $U$ by comparing the photometric properties of that
subset of SNLS SNe~Ia for which some of the $g'r'i'z'$ bands map
conveniently onto rest-frame $UBV$. They estimate the quantity
$\Delta\,U_3$ which is the difference between the $U$ band flux at
maximum light predicted from a suitably-chosen triplet of observed
filters drawn from $g'r'i'z'$ and the actual observed $U$ magnitude. As
the machinery adopted by Astier et al. includes provision for a
color-stretch relation, this is not quite the same test as that
adopted by Jha et al., although the local scatter found in
$\Delta\,U_3$ is comparable (0.12).

\begin{figure}
\plotone{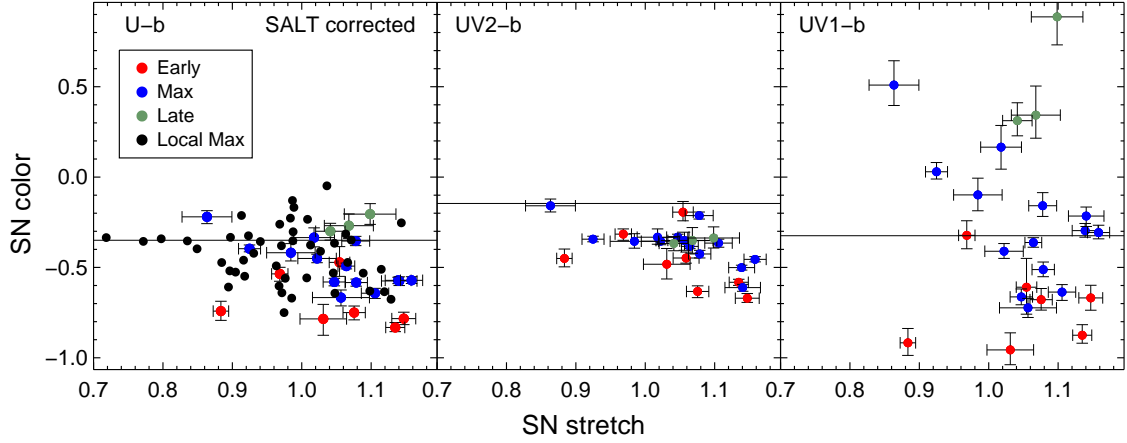}
\caption{Various SN~Ia stretch-color relations, with colors measured
  directly from the SALT color law corrected Keck spectra. The colors
  shown are $U-b$, $UV1-b$ and $UV2-b$ -- see Fig.~\ref{fig:col_lentz}
  for filter definitions.  SNe are color-coded according to the phase
  of the spectrum as defined in Fig.~\ref{fig:redshiftphasehist}. The
  black points in the $U-b$ distribution refers to the $U-b$
  distribution of local SNe~Ia at maximum light from local SN surveys
  \citep[e.g.][]{2006AJ....131..527J}. The error-bars for the SNLS SNe
  include propagated uncertainties from host galaxy subtraction
  ($\S$~\ref{sec:spectr-calibr-host}).\label{fig:col_stretch}}
\end{figure}

We compare our intermediate redshift $U-b$-stretch relation with that
derived from the local sample in Fig.~\ref{fig:col_stretch}.  In this
figure all colors are color-corrected using a SALT color law and the
method of $\S$~\ref{sec:dust}. To derive the rest-frame $U-b$ and $c$
measures for the low-$z$ sample, we re-perform the light curve fits to
the SN photometry using the same fitter used for the SNLS sample
(SiFTO; Conley et al. in prep). Clearly, at maximum light the trends
are very similar.  As discussed in
$\S$~\ref{sec:selection-cfhtls-sne}, the Keck sample (and the SNLS
sample in general) is slightly under-represented in low-stretch SNe~Ia
compared to the local data (Fig.~\ref{fig:samplecomparison}), so the
color-stretch relation found in previous studies is less evident.
Crucially, there is no evidence that the colors derived directly from
the Keck spectra show a larger photometric dispersion.  For events
with stretch $s>$0.85 we find $(U-b)$=-0.452 with a standard deviation
of 0.132 for 15 Keck SNe~Ia c.f. -0.436 (0.167) for 41 local events.
The small difference in mean color probably arises from the different
selection criteria, as Jha et al. are able to include events with
greater extinction. Note there are small differences in the definition
of ``maximum light''; the Keck spectra range over -4 to +4 days which
will add an additional scatter to this sample in the presence of color
evolution with phase.

A similar behavior is seen in the $(UV2-b)$\,-\,stretch relation; a
modest dispersion together with a weak correlation with stretch and a
slight phase dependence (the definitions of UV1 and UV2 are shown in
Fig.\ref{fig:col_lentz}).  However, a marked increase in scatter is
seen in the $(UV1-b)$\,-\,stretch relation (as expected if the UV1
filter is the more sensitive to variations in progenitor composition)
together with a stronger evolution with phase.  Although the UV1
filter is the shortest wavelength probed by the Keck spectra, this
dispersion is considerable even within a phase bin, and is unlikely to
arise from spectrophotometric errors given the tight dispersion
observed in the $UV2-b$ relation.  Fig.~\ref{fig:col_stretch} also
shows weak correlations between color and stretch in the sense that
the high stretch (brighter) events are bluer than those of low stretch
(fainter)
\citep[e.g.][]{1999AJ....118.1766P,2003ApJ...598..102K,2004ApJ...613.1120G}.

Clearly as the SNe~Ia develop they display an increased dispersion in
the wavelength region where metal-dependent features are expected to
be most prominent. This may indicate that the metallicity of the
outer, unburned layers, seen only at early times, is not as important
as expected by \citet{1998ApJ...495..617H} and
\citet{2000ApJ...530..966L}.  Instead line blanketing from iron-peak
elements may play the dominant role in affecting the UV opacity.  This
may affect maximum-light and post-maximum spectra more than for
pre-maximum spectra. At early times the photosphere has not yet
receded into the bulk of the iron-peak elements, and, as time
progresses, $^{56}$Ni synthesized in the explosion decays into
$^{56}$Co and subsequently into $^{56}$Fe. These additional isotopes
may provide the increased line blanketing seen in the spectra.

\subsubsection{Spectroscopic Features}
\label{sec:spectroscopic_features}

Next we consider trends in the spectroscopic features introduced by
\citet{2000ApJ...530..966L} as possible metallicity diagnostics; these
are marked as $\lambda_1$ and $\lambda_2$ in Fig.~\ref{fig:col_lentz}.
The wavelengths of these features have the advantage of being 
independent of any dust correction. If progenitor evolution is largely
driven by metallicity effects, we might expect significant shifts to occur
in these features between the local and high-redshift data.

In fact, a cursory examination of our mean spectra
(Figs.~\ref{fig:mean_max} and ~\ref{fig:mean_early}) shows
that the features do not appear to shift significantly in the mean.
However this could be due to an inadequate or unrepresentative local
data. We have seen that this is a significant limitation.

Accordingly, we measured the wavelength of both features for each of
our spectra individually. For each feature we fit a Gaussian on a
linear continuum background, returning the central wavelength. The
error is estimated by varying the continuum definition either side of
the feature, performing the fit 100 times, and taking the standard
deviation of the resulting central wavelength distribution.

\begin{figure}

\includegraphics[width=3.5in]{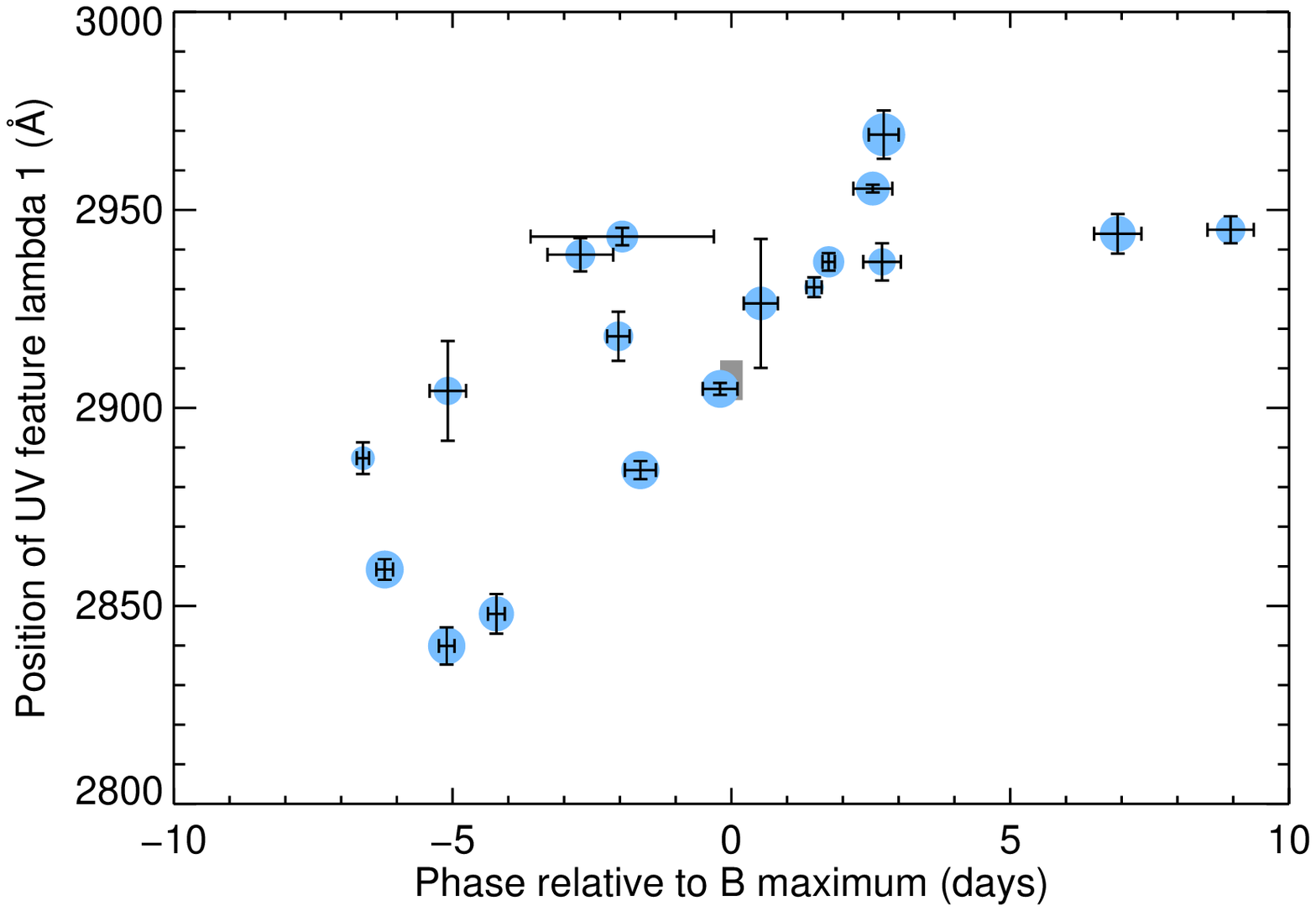}
\includegraphics[width=3.5in]{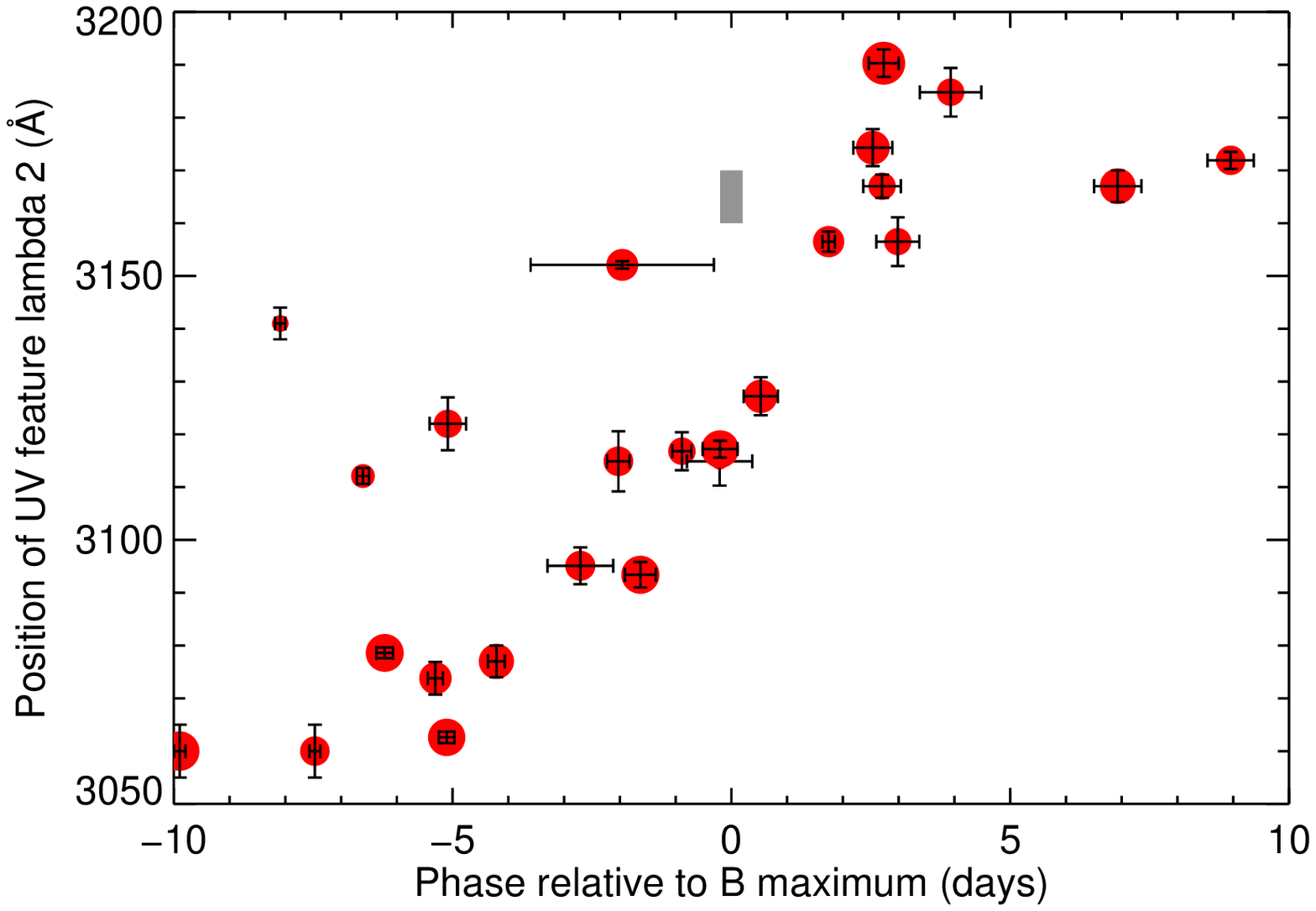}
\caption{Variation of wavelength with phase for the two UV diagnostic
  features marked in Fig.~\ref{fig:col_lentz}; a clear trend is
  apparent. The vertical line refers to the full extent of the
  metallicity variation predicted by
  \citet{2000ApJ...530..966L}.\label{fig:bumps_phase}}
\end{figure}

The precision is typically $\pm$2-5\AA, adequate to detect the shifts
of $\simeq$10\AA\ predicted by \citet{2000ApJ...530..966L}. We found a
much greater SN to SN scatter of $\pm$60\AA, with no obvious trend
with stretch as might be expected from
Fig.~\ref{fig:col_stretch}.  However, as shown in
Fig.~\ref{fig:bumps_phase}, for both features there is a possible
shift with phase in the sense both move gradually to longer wavelength
as the SN expands. Moreover, there is a considerable dispersion at a
particular phase, particularly before maximum light.

\begin{figure}
\plotone{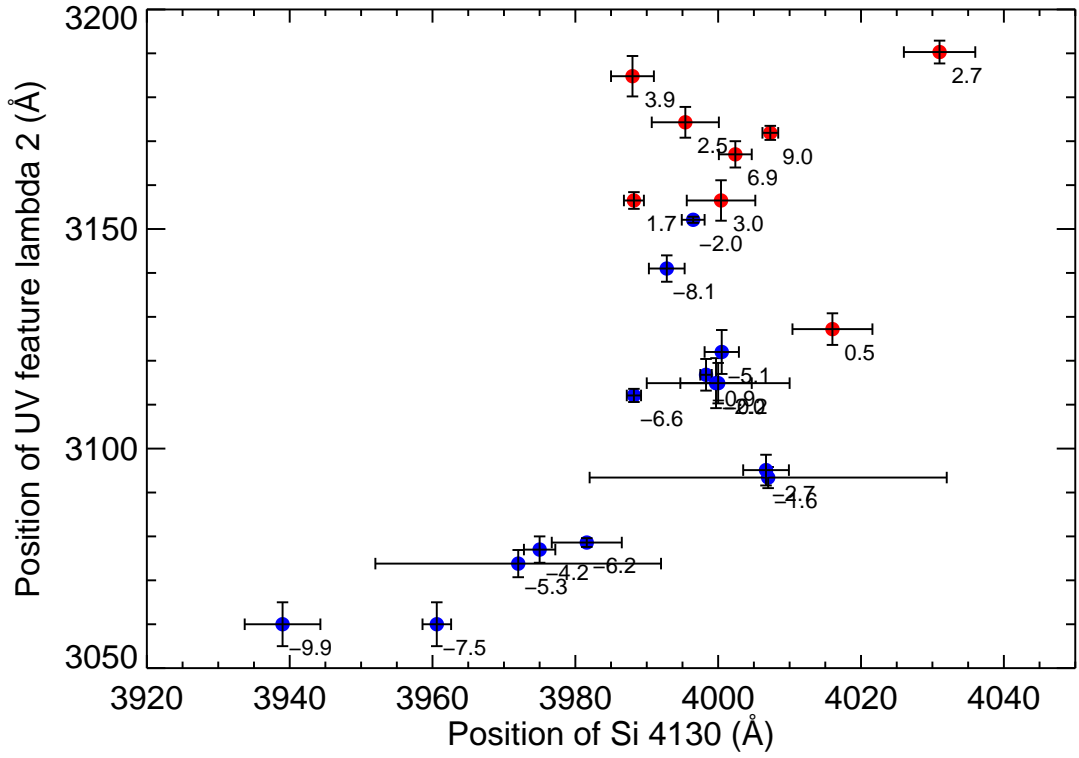}
\caption{Comparison between the wavelength of the \ion{Si}{2} 4130\AA\
  feature commonly used as a measure of the photospheric expansion
  velocity and that of the UV diagnostic $\lambda_2$ (see
  Fig.~\ref{fig:col_lentz} for definition). Spectra taken before
  maximum light are indicated with blue symbols; those after maximum
  light with red symbols. Numbers indicate the phase in rest-frame
  days.\label{fig:siII_lambda2}}
\end{figure}

\begin{figure}
\plotone{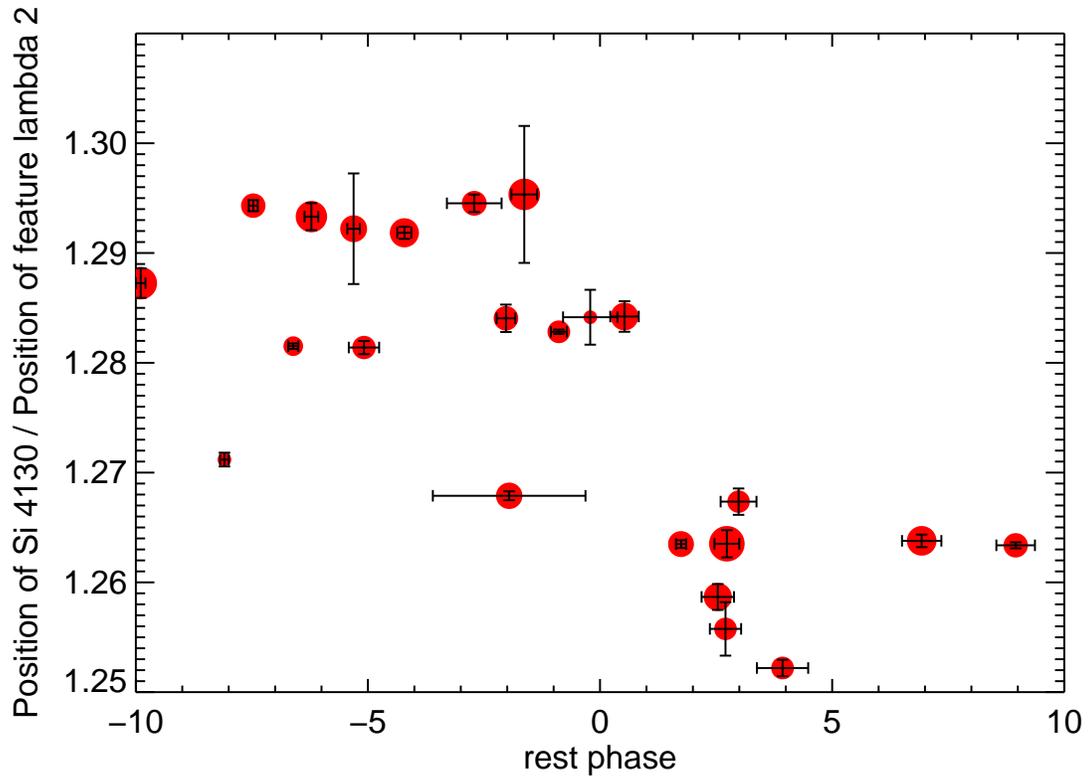}
\caption{Ratio of the wavelength of the \ion{Si}{2} 4130\AA\ feature
  to the UV diagnostic $\lambda_2$ versus phase with data points sized
  according to the stretch. There is a clear discontinuity in the
  behavior after maximum light.\label{fig:ratio_phase}}
\end{figure}

We find that $\lambda_2$ correlates reasonably well with the
wavelength of the \ion{Si}{2} 4130\AA\ - a line commonly used to
measure the photospheric expansion (Fig.~\ref{fig:siII_lambda2}).
Thus, to first order, the scatter seen in $\lambda_2$ at a particular
phase appears to arise from variations in kinetic energy within the
sample. However the phase-dependent trend suggests a further process
is at work. Fig.~\ref{fig:ratio_phase} shows the wavelength ratio
versus phase, coded by stretch, where there is a noticeable shift at
maximum light (corresponding to a redward shift of 50\AA\ in
$\lambda_2$).  This shift presumably arises as a result of the effects
of temperature and ionization changes which are more pronounced in the
UV feature.

In summary, therefore, the above discussion suggests it will be very
difficult to use these wavelength features as diagnostics of
progenitor properties, unless perhaps larger samples are available.
Clearly the analysis in Lentz et al. is oversimplified in several
respects. Their models considered metallicity variations for a fixed
luminosity and epoch, whereas it seems a wide range will be needed to
isolate the effects of metallicity on this part of the spectrum.
Recent work by \citet{2007ApJ...656..661K}, in which a grid of
light curve models that span a large range in the production of
$^{56}$Ni and intermediate mass material (and hence overall kinetic
energy), should provide an excellent starting place to address these
factors in future work.

\subsubsection{Host Galaxy and Stretch Dependences}
\label{sec: host_dependences}

In view of recent work using the SNLS \citep{2006ApJ...648..868S} where
convincing correlations have been found between the properties of
distant SNe~Ia and their host galaxy environment, it is also natural to
question whether the dispersion implied by the
increased scatter in Fig.~\ref{fig:col_stretch} is
dependent on the stellar population and total stellar mass. 

The Keck SN~Ia dataset samples a wide range of passive and
star-forming host galaxies (Fig.~\ref{fig:samplecomparison}) for
which stellar masses and luminosities are available \citep[see][for
details]{2006ApJ...648..868S}. Specific star formation rates and
masses have been measured for the hosts using SED fits to the
extracted colors with the Z-PEG photometric redshift code
\citep{2002A&A...386..446L}.

\begin{figure}
\plotone{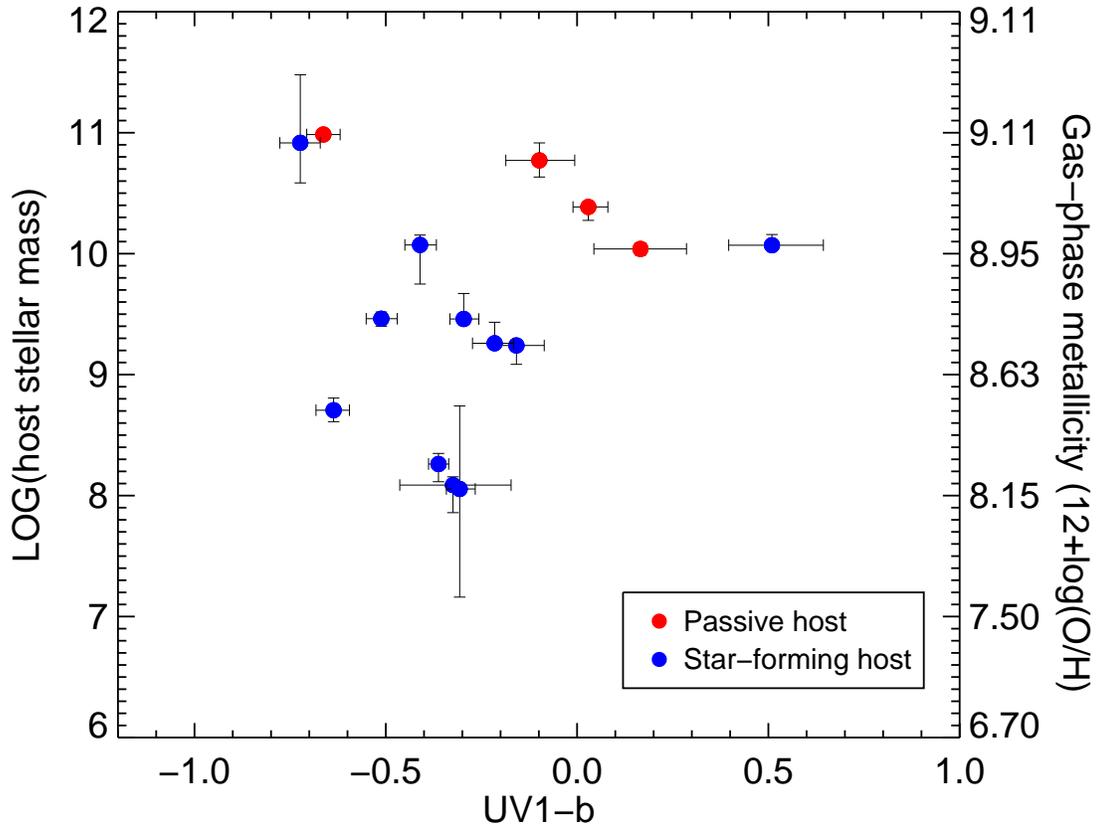}
\caption{The variation of the Keck SN~Ia $UV1-b$ spectral color as a
  function of the stellar mass of the SN host galaxy color coded
  according to the specific star formation rate. The division is made
  at a rate of 10$^{-12} M_{\odot}$ yr$^{-1}$ per unit stellar mass.
  The color error-bars include propagated uncertainties from host
  galaxy subtraction. The conversion from stellar mass to gas-phase
  metallicity using the relation of \citet{2004ApJ...613..898T} is
  also shown.\label{fig:col_mass}}
\end{figure}

In Fig.~\ref{fig:col_mass} we plot the color-corrected $UV1-b$ color
against the host-galaxy stellar mass coding the data points according
to whether the hosts are primarily passive or active. The
discriminating specific star-formation rate was chosen to be 10$^{-12}$ M$_{\odot}$
yr$^{-1}$ per unit stellar mass (Fig.~\ref{fig:samplecomparison}).
Gas phase metallicities in star forming galaxies are well-known to
correlate with galaxy mass
\citep{2004ApJ...613..898T,2003ApJ...599.1006K} and recent studies
have revealed these correlations were already in place at
$z\simeq$1-1.5 \citep{2005ApJ...635.1006S}. Over the 3~dex range in
stellar mass seen in our host galaxies, we would expect a change in
metallicity of over 1~dex.

If, as Sullivan et al. surmise, the higher SN~Ia rate in active hosts
indicate the bulk of those SNe formed relatively recently, one might
expect a tighter or different correlation between metallicity and
galaxy mass than for those in passive galaxies. Unfortunately,
Fig.~\ref{fig:col_mass} reveals an unavoidable difficulty with the
modest Keck sample; our four passive galaxies are considerably more
massive than most of the star-forming galaxies. Thus it is not
straightforward to separate the effects of galaxy mass and specific
star formation rate in understanding the implied metallicity
dispersion. Nonetheless, to the extent we can examine any
relationships, it seems there are much stronger trends between $UV1-b$
and inferred metallicity for SNe~Ia in star-forming hosts than for
those with passive hosts. At face value this would imply the most
metal-rich progenitors have {\it less} UV blanketing. Clearly a larger
sample will be needed to verify these trends. It may not be
straightforward to sample lower-mass passive hosts, given fewer
passive hosts are expected at high redshift in the hypothesis proposed
by Sullivan et al..

A more informative comparison of the UV trends can be obtained by
examining the spectra for those SNe~Ia with high and low stretch.
Although the host galaxy spectral class may contain events with a
range of stretch (c.f. fig.~11 of \citet{2006ApJ...648..868S}), such a
categorization of events may be a more direct way of examining the
spectral properties of events in old and young stellar populations
\citep[see also][]{2007bronder}.

\begin{figure}
\includegraphics[width=3.5in]{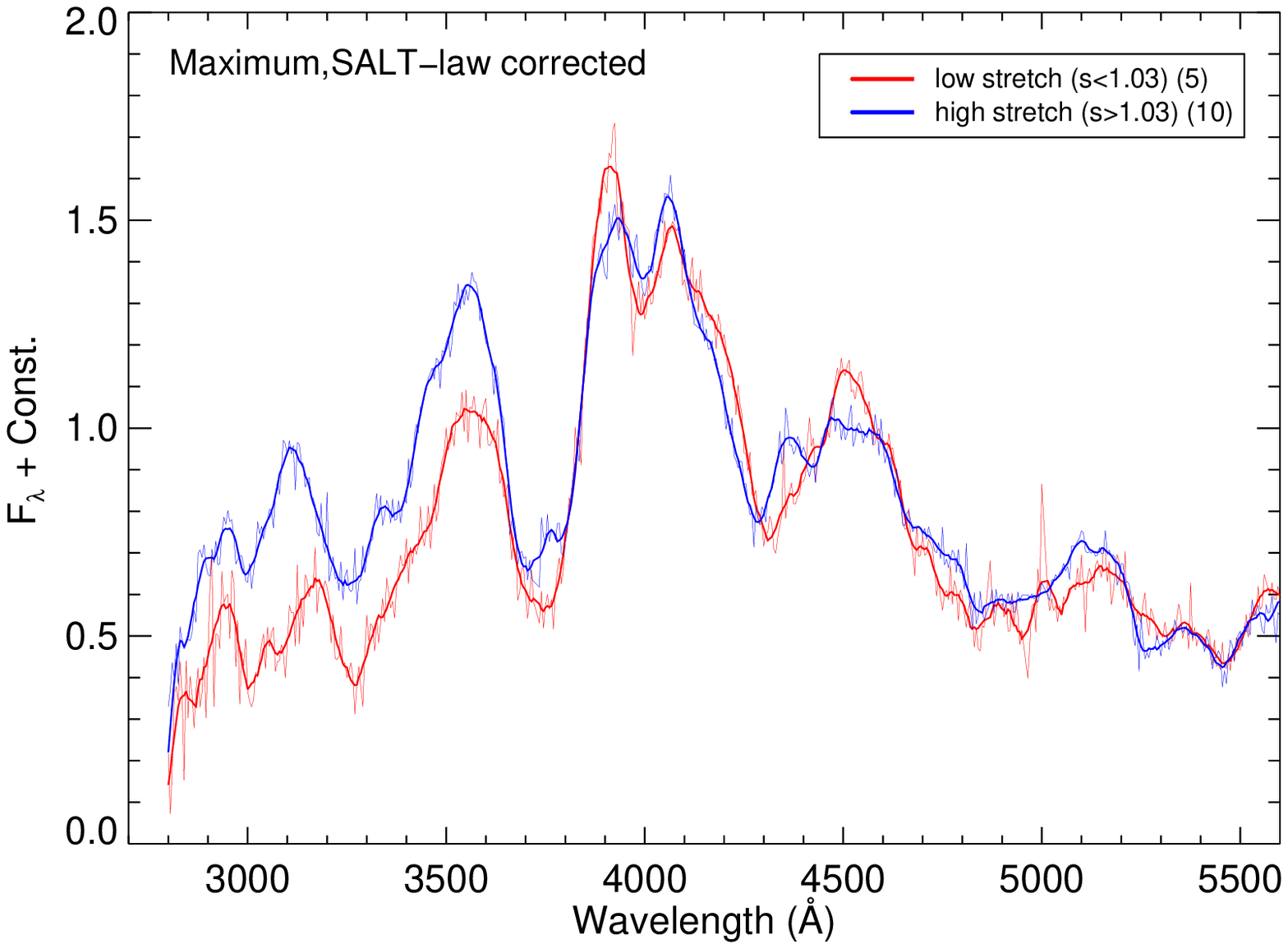}
\includegraphics[width=3.5in]{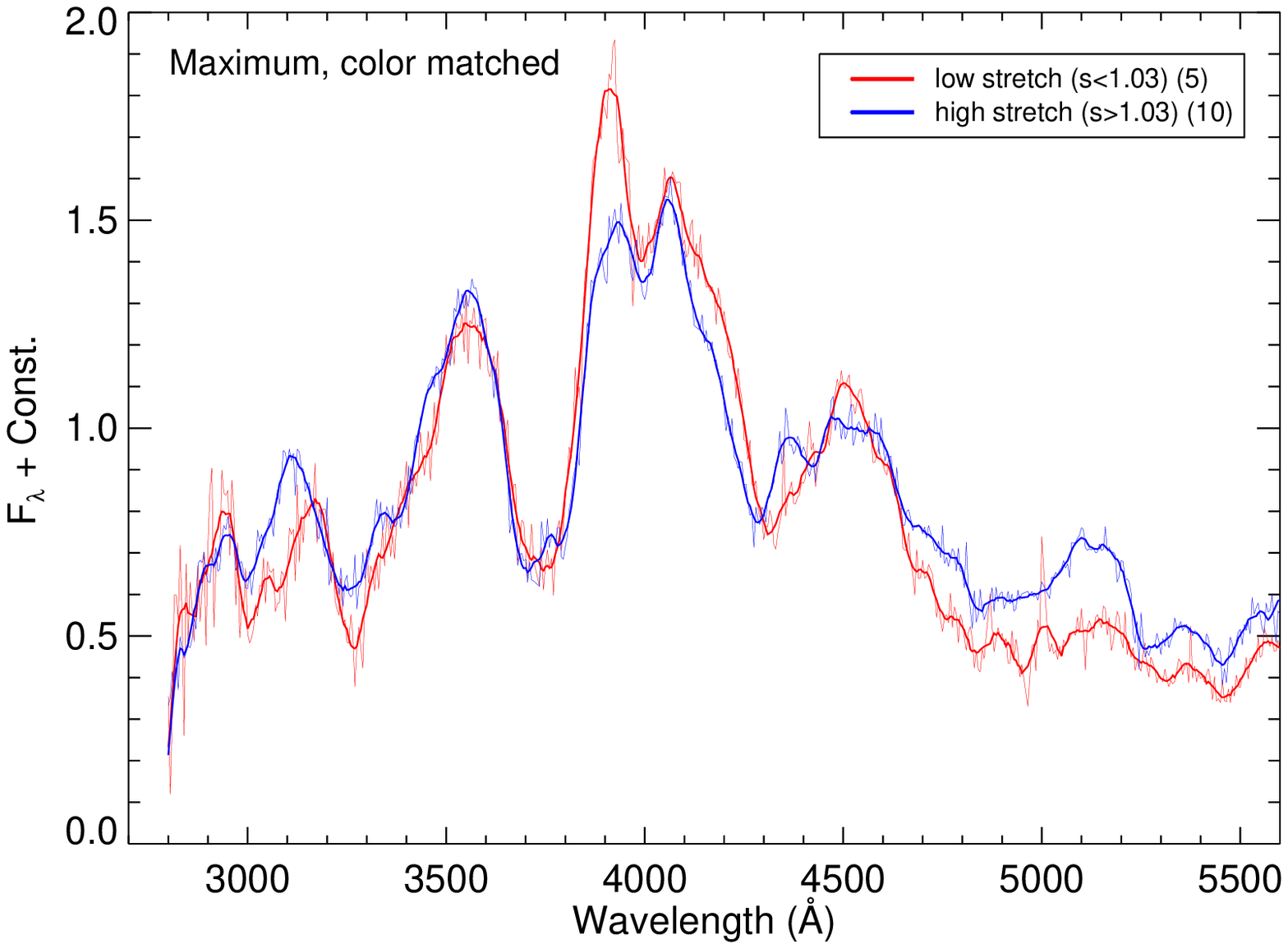}
\caption{The mean SALT-law color-corrected (left) and color-matched
  (right) Keck spectra for SNe~Ia split according to the SN light
  curve stretch, $s$.  The red spectrum refers to events with
  $s\leq1.03$, the blue spectrum to those with
  $s>1.03$.\label{fig:mean_max_s}}
\end{figure}

Fig.~\ref{fig:mean_max_s} shows the mean maximum-light SALT-law
color-corrected spectra for those events with stretch s$>$1.03 and
s$\leq$1.03. We also show the mean spectra split by $s$ in the same
way when the spectra are color-matched (rather than color-corrected)
to have the same $U-b$ color. These comparisons confirm in detail the
photometric trend noted in the rightmost panel of
Fig.~\ref{fig:col_stretch} -- a strong and systematic difference in
the far UV maximum-light spectra of low and high-$s$ SNe~Ia. 

As stretch (or light curve shape) maps closely onto both
star-formation activity in the host galaxy \citep{2006ApJ...648..868S}
and the morphology of the host galaxy
\citep[e.g.][]{2000AJ....120.1479H}, this comparison provides a clear
example of the physical differences between SNe~Ia in star forming and
passive hosts and hence SNe~Ia drawn from young and old stellar
populations.

Clearly the mean spectrum of a SN~Ia in a passive host with an old
stellar population has a depressed far-UV continuum compared to one in
a more active host with a younger stellar population even following
color corrections based on optical colors. A broadly similar trend is
observed by \citet{2007A&A...466...11G}. This is in the opposite sense
than that expected from dust extinction, where the higher-stretch
SNe~Ia residing in spiral galaxies should suffer more extinction and
therefore appear redder in the UV. Furthermore, the far-UV colors
appear more sensitive to stretch than the optical colors.

Fig.~\ref{fig:mean_max_s} also reveals interesting differences between
the optical spectral regions of high and low stretch supernovae.  The
majority of these differences originate due to the underlying
temperature difference, with the high stretch supernovae showing
stronger high excitation lines and ones from doubly-ionized elements.
At $\simeq$4400 \AA, a line due to \ion{Si}{3} 4552 \AA\ is found in
the high stretch supernovae but is absent in the lower stretch
examples. In addition, the 3700 \AA\ Ca+Si feature is narrower and
shows a greater degree of splitting into the separate Ca and Si
components in the high stretch composite spectrum. Here one can also
clearly see the difference between $\cal{R}$(\ion{Ca}{2}) as defined
in \citet{nugseq95} where the ratio of the emission features on either
side of this trough are greater in the low stretch composite. Finally,
the \ion{Si}{2} feature at 4130\AA\ is shallower in high-stretch SNe
consistent with trends identified by \citet{2007bronder}.

\section{Discussion}
\label{sec:discussion}

One of the original motivations for this study was to explore both
evolution in the mean properties of SNe~Ia over 0$<z<$0.5 and the
dispersion in their UV spectra. We have seen remarkably little
evidence of evolution in the mean spectroscopic properties over this
interval, confirming, with higher precision, earlier suggestions based
largely on lower-S/N data.  The primary limitation in this aspect of
our paper is the paucity of high quality UV data at low redshift.

Most of the earlier constraints on the possible evolution of SNe~Ia
were largely based on consideration of photometric measures. For
example, \citet{2004ApJ...607..665R} derived the rest-frame $U-B$ and
$B-V$ distributions of a heterogeneous sample of SNe~Ia at maximum
light over z$<$1.5 as a proxy for the more precise constraints
possible with spectroscopic data.  They ruled out any color evolution
greater than 0.02~mag in the mean in $U-B$.

Spectroscopic data is substantially more precise in its tracking of
evolution and dispersion than broad-band photometry for two reasons.
Foremost, detailed differences in key line diagnostics are lost in
photometric data. Secondly, no assumptions need be made about
$k$-corrections in gathering and comparing data over a broad redshift
range.

At the time this Keck survey was underway, there was no comprehensive
spectroscopic dataset against which our UV data could be compared,
other than the mean template published by \citet{2002PASP..114..803N}
and discussed in $\S$~\ref{sec:local-data}. However, very recently, as
a result of a survey of high-redshift SNe~Ia with the ACS grism
onboard Hubble Space Telescope, \citet{2007ApJ...659...98R} have
published the mean rest-frame UV spectrum of 13 SNe~Ia more distant
($z>1$) than those studied here; the average redshift of this sample
is $\overline{z}$=1.3. In this study, used to demonstrate the presence
of dark energy at $z>$1, it is claimed that the sample-averaged
spectral energy distributions observed at $z>$1 are consistent with
that observed locally and that any spectral evolution is still
undetected.  However, no quantitative statement is made to support
this claim.

We believe that the spectral comparison undertaken by
\citet{2007ApJ...659...98R} is less precise than that undertaken in
this study, and hence less valuable as a means of justifying the
continued use of local relations in constructing SN~Ia Hubble
diagrams.  Foremost, the ACS grism data samples SNe~Ia over a much
wider range of phase than the comparisons undertaken here.  Although
the individual spectra comprising the mean $z$=1.3 ACS spectrum are
not tabulated by \citet{2007ApJ...659...98R}, their phases range over
at least 15-20 days and almost all are post-maximum light, compared
with the narrow phase range maximum-light (-4 to +4 days) and
pre-maximum light ($<$-4 days) comparisons presented here.  A second
consideration is the inevitable poor signal-to-noise ratio of these
very high redshift spectra which precludes detailed comparisons.

As we have shown in Figs.~\ref{fig:mean_max} and
~\ref{fig:mean_early}, the local template
\citep{2002PASP..114..803N} is a poor basis from which to make such
comparisons. More representative datasets are needed for reliable
claims. Internally within our own data, Fig.~\ref{fig:mean_max_z}
reveals little evolution although the redshift baseline is small. It
will be important to examine the case for evolution from 0.2$<z<$1.5
by combining in a consistent manner both the
\citet{2007ApJ...659...98R} dataset with that presented here, as well
as with higher-redshift SNLS SNe~Ia (out to $z\sim1$) observed during
the routine survey spectroscopic screening
\citep[e.g.][]{2007bronder}.

The second significant finding in this study is the marked increase in
the dispersion among our SNe~Ia shortward of 3300\AA\
(Fig.~\ref{fig:col_stretch}). Although the scatter in $U$ is
comparable to that seen in local data, it increases significantly at
shorter wavelengths.  Even allowing for differing amounts of dust
extinction within each host galaxy, this amounts to more than a factor
of 2 variation in continuum flux at maximum light.  Although
theoretical models predict a strong sensitivity to metallicity
variations at this wavelength, our variations are also considerably
larger than those predicted.  We have demonstrated that a color
correction based on the $B-V$ color of the SN and either a CCM Milky
Way extinction law or a SALT color law, can only marginally reduce, and
not eliminate, this UV dispersion.

We have found it hard to isolate the physical causes of this
significant UV dispersion. We confirm earlier work at longer
wavelengths ($\lambda \simeq$3500-4000\AA) that shows that stretch
(or equivalently host galaxy class) is partially responsible. However
at the shortest wavelengths ($\lambda\simeq$3000-3300\AA\
corresponding to our UV1 diagnostic), additional effects are clearly
important that are not accounted for by the color-correction
techniques in use in current cosmological programs. In addition to the
newly-found intrinsic scatter at short wavelengths, new trends with
phase are also seen in the wavelengths of diagnostic features in this
region.

Redward of 4000 \AA\ , corrections for stretch or color work well in
normalizing SNe Ia, however blueward of 4000A, significant scatter
remains even after such corrections are made.  This may be related to
a change in the dominant source of opacity in SNe\,Ia.  Redward of
3500-4000 \AA\, electron scattering opacity dominates, but at UV
wavelengths a forest of overlapping lines is the dominant source of
opacity (see Fig. 1 of \cite{2000ARA&A...38..191H}).
Electron scattering is a continuous process involving well-understood
physics, but line opacity depends sensitively on abundances,
ionization states, and possibly non-LTE effects.

\begin{figure}
\centering
\includegraphics[width=5.5in,angle=270]{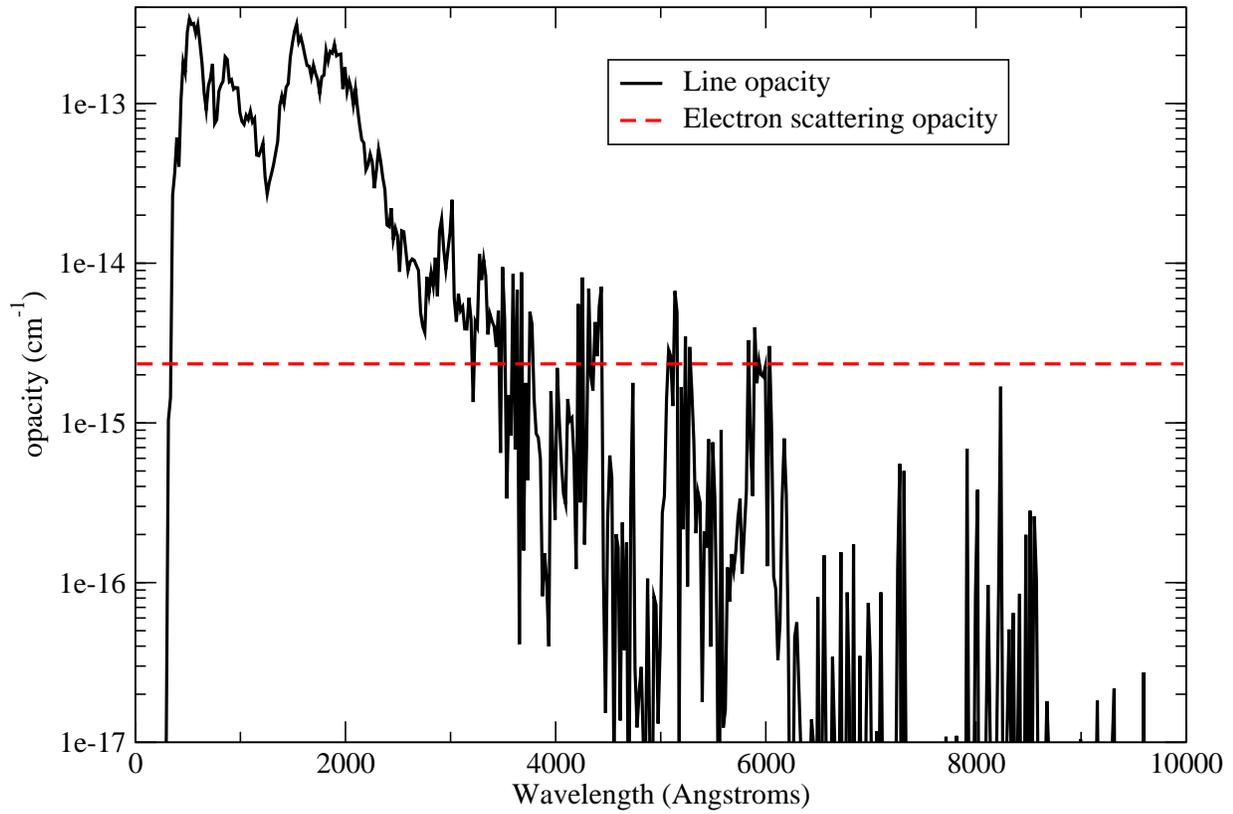}
\caption{Comparison of the wavelength-dependent line and electron
  scattering opacity for a typical model in
  \citet{2007ApJ...656..661K} at peak SN\,Ia brightness. The data
  refers to a depth of 7000 km\,s$^{-1}$. Note the drop in line
  opacity with respect to the electron scattering opacity near 4000
  \AA\ . This behavior makes the emergent UV flux highly sensitive to
  changes in the line opacity whereas the optical and near-IR spectral
  regions are largely dominated by electron scattering
  opacity.\label{fig:opac_max}}
\end{figure}

Fig. ~\ref{fig:opac_max} illustrates this point by comparing the wavelength 
dependence of the line and electron scattering opacity at maximum light 
for a model by \cite{2007ApJ...656..661K} that provides a good match 
for a normal SN~Ia (Kason, priv. commun.). The data in question refers 
to that at a depth of 7000 km\,s$^{-1}$ where intermediate mass and 
Fe peak material are well-mixed. A common feature in these models is the 
drop in line opacity compared to the electron scattering opacity near 
4000 \AA\ as seen here. Thus it is understandable that the emerging 
UV flux is highly susceptible to changes in the line opacity (due to initial 
conditions and/or material synthesized during the explosion) while the 
optical and near-IR spectral behavior are dominated by electron 
scattering opacity at this phase.

What are the possible consequences of the above variations in terms of
the use of SNe~Ia as probes of the expansion history? In the highest
redshift surveys, including those proposed with future facilities,
cross-color $k$-corrections are needed to estimate rest-frame light
curves from the observations, typically undertaken in the far-red and
near-infrared for $z>$1.5 \citep{2004ApJ...607..665R,
  2007ApJ...659...98R}.  This necessitates the adoption of a suitable
template whose SED is reliable in the 3000-4000\AA\ region. At the
most fundamental level, the dispersion in our UV spectra will
contribute a statistical uncertainty at the 0.05-0.1 magnitude level
depending on how precisely particular observed filters match to the
chosen rest-frame bandpasses.

A more worrying bias would follow the adoption of an incorrect
template.  Possibly the most significant finding in our work is the
demonstration that the mean UV spectrum is different for high and low
stretch SNe~Ia (Fig.~\ref{fig:mean_max_s}) and thus, presumably, for
those that occur in passive and actively star-forming galaxies
\citep{2006ApJ...648..868S}.  Adopting a single template would lead to
a systematic bias which could become increasingly serious at high
redshift where the mix between the two populations changes and
high-stretch SNe, typically originating from younger progenitor
systems, become increasingly common. Recently,
\citet{2007astro.ph..1912H} have demonstrated the existence of an
increased fraction of high-stretch SNe~Ia at higher redshifts above
and beyond the selection effect that high stretch, brighter SNe~Ia are
easier to detect.

\begin{figure}
\plotone{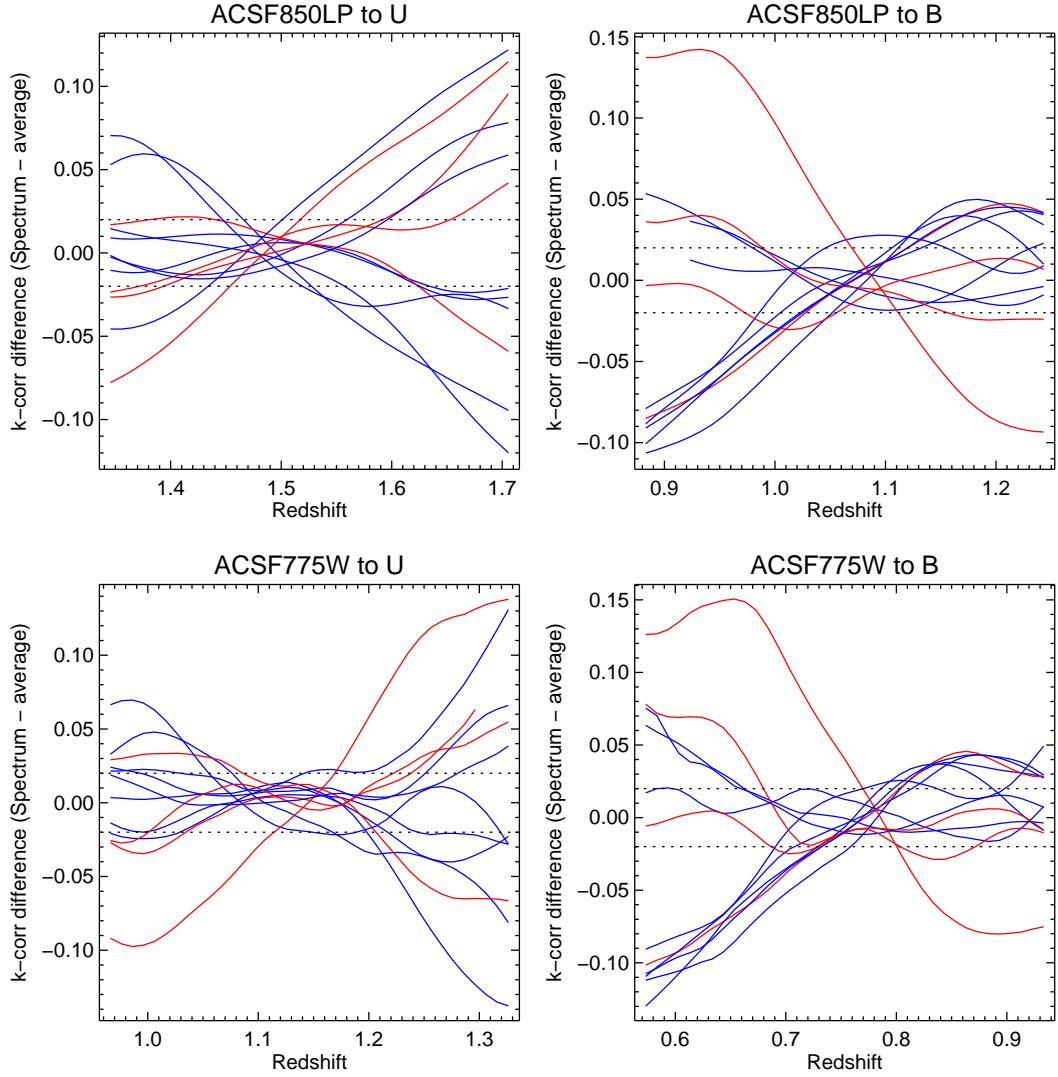}
\caption{Redshift dependence of various maximum light cross-filter
  $k$-corrections differenced to that based on the local template.
  Lines represent deviations observed for individual Keck spectra
  color-coded according to stretch: red lines refer to events with
  $s\leq1.03$, while blue to those with $s>1.03$. The dashed lines
  indicates the approximate precision necessary to secure an equation
  of state parameter $w$ to a precision of 5\% using $z>$1 SNe\,Ia
  (see text for discussion).\label{fig:kcorrections}}
\end{figure}

Using our individual maximum light spectra, we can evaluate the impact of
the UV dispersion on sample cross-color $k$-corrections, addressing
both the statistical error arising from the intrinsic scatter and the
systematic difference arising from the stretch-based spectral
differences we have found.  Fig.~\ref{fig:kcorrections} shows the
result. 

For a typical future dark energy experiment, based on securing
the equation of state parameter $w$ to 5\% using $z>$1 SNe\,Ia,
photometric corrections of better than $\pm$0.02 mag are required
\footnote{e.g. White Papers submitted to the 
US Dark Energy Task Force Committee:
{\rm http://www.nsf.gov/mps/ast/detf.jsp}}. This requirement is
indicated on Fig.~\ref{fig:kcorrections}. 

In terms of the {\em statistical} error, the observed dispersion 
is typically $\pm$0.05-0.10 mag, larger than that required by a factor of
several. However, such a dispersion, if randomly distributed, need not 
present a fundamental obstacle to progress with a survey spanning a
wide redshift range utilizing a large number of SNe\, Ia. As 
Fig.~\ref{fig:kcorrections} shows, at certain redshifts where the rest-frame 
and observed filters closely match, the dispersion has negligible effect 
and SNe\,Ia at these redshifts could be more heavily weighted.

The more worrying trend, particularly given the recent demonstration
of stretch bias by \citet{2007astro.ph..1912H}, is the likelihood of a 
{\em systematic} error introduced by adopting an incorrect template for 
the redshifts where the rest-frame and observed filters do not overlap. 
Fig.~\ref{fig:kcorrections} shows the potential of this error via a 
comparison of the dispersion independently for high and low stretch events.
The differences between these two categories are comparably large and 
indicate the importance of securing a physical understanding of the 
UV variations seen in our survey.

\section{Conclusions}
\label{sec: conclusions}

We summarize our findings as follows:

\begin{enumerate}

\item{} We have secured high signal-to-noise ratio Keck spectra for a
  sample of 36 intermediate redshift SNe~Ia, observed at various
  phases, spanning the redshift range 0.15$<z<$0.7, and drawn from the
  Supernova Legacy Survey (SNLS). We demonstrate via inspection of the
  SN properties that our Keck sample is a reasonably fair subset of
  the larger sample of distant SNe~Ia being studied by the SNLS.

\item{} We develop a new method for removing host galaxy contamination
  from our spectra based on measures of the galaxy and SN photometry.
  These refinements to traditional spectral reduction techniques allow
  us to achieve host-galaxy subtracted and flux-calibrated rest-frame
  spectra of high quality, extending down to rest-frame wavelengths of
  2900\AA.

\item{} Although no strong evidence is found for spectral evolution in
  the mean early-phase and maximum light spectra, when compared to
  local data, such evolutionary tests are hampered by the paucity of
  quality data at low redshift and a significant scatter in the
  spectra shortward of 4000\AA. We find no evidence for evolution
  internal to our data.  We argue that the well-used local UV spectral
  template \citep{2002PASP..114..803N} is likely to be less
  representative than the mean spectrum compiled from the Keck data
  which we tabulate with the measured dispersion for use in future
  cosmological applications.

\item{} Our principal finding is a large scatter from one SN to the
  next in the rest-frame UV spectrum even after standard dust
  corrections are made. By constructing various photometric bandpasses
  that avoid uncertainties arising from differential $k$-corrections
  associated with the range of redshifts in our sample, we show that
  while we can reproduce the stretch-dependent trends seen locally at
  3500-4000\AA, the scatter at 3000-3400\AA\ is 3-5 times larger.

\item{} Although progenitor metallicity may drive some of the trends
  seen in the Keck data, the UV variations are much larger than in
  contemporary models which span the expected metallicity range.
  Moreover, the UV spectrum also changes with phase in a manner which
  is not consistent with models. We conclude there are significant
  variations in the UV properties of SNe~Ia which are not accounted for
  by either the presently-employed empirical trends or the available
  SN~Ia models.

\item{} As an illustration of the importance of understanding these
  new results, we calculate the error arising from the use of a single
  UV spectral template for calculating the cross-color $k$ correction,
  a correction essential for constructing the SN~Ia Hubble diagram as
  a probe of the expansion history. The dispersion arising from our UV
  spectra, if not randomly distributed along the Hubble diagram, presents 
  an uncertainty 2-3 times larger than than would be
  necessary for recovering the equation of state parameter $w$ to 5\%
  using SNe~Ia at $z\simeq$1. We conclude that further detailed
  studies are essential if SNe~Ia are to be useful for precision
  measures of dark energy.

\end{enumerate}

\acknowledgments

The spectroscopic data presented herein were obtained at the W.M. Keck
Observatory, which is operated as a scientific partnership among the
California Institute of Technology, the University of California and
the National Aeronautics and Space Administration. The Observatory was
made possible by the generous financial support of the W.M. Keck
Foundation. The authors wish to recognize and acknowledge the very
significant cultural role and reverence that the summit of Mauna Kea
has always had within the indigenous Hawaiian community.  We are most
fortunate to have the opportunity to conduct observations from this
mountain. Additional observations were obtained with
MegaPrime/MegaCam, a joint project of CFHT and CEA/DAPNIA, at the
Canada-France-Hawaii Telescope which is operated by the National
Research Council (NRC) of Canada, the Institut National des Sciences de
l'Univers of the Centre National de la Recherche Scientifique (CNRS)
of France, and the University of Hawaii. Canadian collaboration
members acknowledge support from NSERC and CIAR; French collaboration
members from CNRS/IN2P3, CNRS/INSU and CEA. PEN acknowledges support
from the US Department of Energy Scientific Discovery through Advanced
Computing program under contract DE-FG02-06ER06-04. A.G. acknowledges
support by NASA through Hubble Fellowship grant \#HST-HF-01158.01-A
awarded by STScI, which is operated by AURA, Inc., for NASA, under
contract NAS 5-26555.  RSE acknowledges financial support from DOE
under contract DE-FG02-04ER41316. This research used resources of the
National Energy Research Scientific Computing Center, which is
supported by the Office of Science of the U.S. Department of Energy
under Contract No.  DE-AC03-76SF00098. We thank them for a generous
allocation of computing time. We acknowledge useful discussions with
Adam Riess, Mario Livio, and Ken Nomoto.

{\it Facilities:} \facility{Keck:I}, \facility{CFHT}.

\clearpage

\begin{deluxetable}{lllccccll}
\tablecaption{Distant Supernova Sample}
\tabletypesize{\scriptsize}
\tablehead{\colhead{SN Name$^a$} & \colhead{RA} & \colhead{DEC} & \colhead{Redshift} & \colhead{Phase} & \colhead{Stretch} & \colhead{Nearest $g'$} & \colhead{\%} & \colhead{Redshift}\\ \colhead{} & \colhead{(J2000)} & \colhead{(J2000)} & \colhead{} & \colhead{(Days)$^b$} & \colhead{} & \colhead{phot (Days)$^b$} & \colhead{Inc.$^c$} & \colhead{Source}}
\startdata
03D1au & 02:24:10.38 & -04:02:14.9 & 0.5043\,$\pm$\,0.0005 &  -1.6\,$\pm$\,0.3 & 1.14\,$\pm$\,0.02 &  +0.6 &  432 & Galaxy\\ 
03D1aw & 02:24:14.78 & -04:31:01.6 & 0.582\,$\pm$\,0.005 &  +2.4\,$\pm$\,0.3 & 1.08\,$\pm$\,0.02 &  +0.6 &  856 & Galaxy\\ 
03D1co & 02:26:16.23 & -04:56:05.7 & 0.679\,$\pm$\,0.001 &  +7.1\,$\pm$\,0.5 & 1.07\,$\pm$\,0.04 &  +0.7 &  321 & Galaxy\\ 
03D1dj & 02:26:19.08 & -04:07:09.3 & 0.40\,$\pm$\,0.01 &  -1.8\,$\pm$\,2.3 & 1.25\,$\pm$\,0.61 &  +0.1 & 4199 & Template Fit\\ 
03D3af & 14:21:14.92 & +52:32:15.3 & 0.5320\,$\pm$\,0.0005 &  +2.8\,$\pm$\,0.4 & 1.02\,$\pm$\,0.03 &  +3.8 &  368 & Galaxy\\ 
03D3aw\tablenotemark{e} & 14:20:53.61 & +52:36:20.6 & 0.449\,$\pm$\,0.001 &  -0.8\,$\pm$\,0.2 & 0.94\,$\pm$\,0.03 & -13.7 &   52 & Galaxy\\ 
03D3ay\tablenotemark{e} & 14:17:58.43 & +52:28:57.4 & 0.3709\,$\pm$\,0.0003 &  -1.0\,$\pm$\,0.2 & 1.00\,$\pm$\,0.03 & -14.6 &   72 & Galaxy\\ 
03D3ba\tablenotemark{e} & 14:16:33.44 & +52:20:32.1 & 0.2912\,$\pm$\,0.0003 & +12.8\,$\pm$\,0.4 & 1.04\,$\pm$\,0.03 & -15.5 &  259 & Galaxy\\ 
03D3bb\tablenotemark{f} & 14:16:18.78 & +52:14:55.3 & 0.2437\,$\pm$\,0.0003 &  +2.2\,$\pm$\,0.3 & 1.18\,$\pm$\,0.03 & -16.2 & 2323 & Galaxy\\ 
03D3bh\tablenotemark{e} & 14:21:35.89 & +52:31:37.6 & 0.2486\,$\pm$\,0.0003 &  -3.5\,$\pm$\,0.2 & 1.02\,$\pm$\,0.03 & -16.1 &  140 & Galaxy\\ 
03D3bl & 14:19:55.90 & +53:05:51.0 & 0.3553\,$\pm$\,0.0005 &  +3.7\,$\pm$\,0.5 & 0.98\,$\pm$\,0.03 &  -0.1 &  270 & Galaxy\\ 
03D3cc & 14:19:45.25 & +52:32:25.3 & 0.4627\,$\pm$\,0.0003 &  +8.5\,$\pm$\,0.4 & 1.04\,$\pm$\,0.02 &  -0.1 &  204 & Galaxy\\ 
03D3cd & 14:18:39.95 & +52:36:43.8 & 0.4607\,$\pm$\,0.0005 &  -5.6\,$\pm$\,0.1 & 1.15\,$\pm$\,0.02 &  -0.1 &  625 & Galaxy\\ 
03D4ag & 22:14:45.79 & -17:44:23.0 & 0.2847\,$\pm$\,0.0003 &  -5.0\,$\pm$\,0.1 & 1.08\,$\pm$\,0.02 &  -1.6 &  141 & Galaxy\\ 
03D4cj\tablenotemark{f} & 22:16:06.66 & -17:42:16.7 & 0.27\,$\pm$\,0.01 &  -7.4\,$\pm$\,0.1 & 1.09\,$\pm$\,0.01 &  +1.5 & 5000 & Template Fit\\ 
03D4dh & 22:17:31.04 & -17:37:46.9 & 0.6268\,$\pm$\,0.0004 &  +0.5\,$\pm$\,0.3 & 1.08\,$\pm$\,0.02 &  +0.6 &  444 & Galaxy\\ 
03D4gl\tablenotemark{e} & 22:14:44.17 & -17:31:44.4 & 0.571\,$\pm$\,0.001 &  +7.7\,$\pm$\,0.5 & 1.00\,$\pm$\,0.03 & +14.6 &  347 & Galaxy\\ 
04D1hd & 02:26:08.85 & -04:06:35.2 & 0.3688\,$\pm$\,0.0004 &  +1.6\,$\pm$\,0.1 & 1.06\,$\pm$\,0.01 &  +1.4 & 1998 & Galaxy\\ 
04D1jg & 02:26:12.56 & -04:08:05.3 & 0.5842\,$\pm$\,0.0005 &  -2.0\,$\pm$\,0.2 & 1.05\,$\pm$\,0.02 &  +1.3 &  128 & Galaxy\\ 
04D1oh\tablenotemark{d} & 02:25:02.37 & -04:14:10.5 & 0.59\,$\pm$\,0.01 &  -6.0\,$\pm$\,0.2 & 1.03\,$\pm$\,0.02 &  +0.1 &  512 & Template Fit\\ 
04D1rh & 02:27:47.16 & -04:15:13.6 & 0.4349\,$\pm$\,0.0005 &  +2.6\,$\pm$\,0.3 & 1.02\,$\pm$\,0.03 &  -0.1 &  393 & Galaxy\\ 
04D1sk & 02:24:22.77 & -04:21:13.3 & 0.6634\,$\pm$\,0.0005 &  -0.4\,$\pm$\,0.6 & 0.86\,$\pm$\,0.04 &  +0.0 &  139 & Galaxy\\ 
04D2gc & 10:01:39.26 & +01:52:59.5 & 0.5216\,$\pm$\,0.0005 &  -0.2\,$\pm$\,0.3 & 1.14\,$\pm$\,0.03 &  -0.9 &  367 & Galaxy\\ 
04D2kr & 10:00:37.32 & +01:42:43.1 & 0.7441\,$\pm$\,0.0005 &  -2.5\,$\pm$\,0.6 & 1.06\,$\pm$\,0.04 &  +0.6 &   91 & Galaxy\\ 
04D3cp\tablenotemark{d} & 14:20:23.95 & +52:49:15.5 & 0.83\,$\pm$\,0.02 &  +5.2\,$\pm$\,0.4 & 1.07\,$\pm$\,0.02 &  +0.5 &  557 & Galaxy\\ 
04D3ez & 14:19:07.91 & +53:04:18.8 & 0.2630\,$\pm$\,0.0005 &  +1.6\,$\pm$\,0.1 & 0.92\,$\pm$\,0.02 &  -0.4 &   78 & Galaxy\\ 
04D3fk & 14:18:26.21 & +52:31:42.7 & 0.3578\,$\pm$\,0.0007 &  -6.9\,$\pm$\,0.1 & 0.97\,$\pm$\,0.01 &  -0.4 &  136 & Galaxy\\ 
04D4in & 22:15:08.58 & -17:15:39.8 & 0.5160\,$\pm$\,0.0005 &  -4.6\,$\pm$\,0.1 & 1.13\,$\pm$\,0.01 &  +2.0 &  443 & Galaxy\\ 
04D4jr & 22:14:14.33 & -17:21:00.9 & 0.482\,$\pm$\,0.007 &  -0.2\,$\pm$\,0.2 & 1.16\,$\pm$\,0.02 &  +2.7 & 5000 & Template Fit\\ 
05D1hk\tablenotemark{f} & 02:24:39.16 & -04:38:03.0 & 0.2631\,$\pm$\,0.0002 &  -8.6\,$\pm$\,0.1 & 1.16\,$\pm$\,0.01 &  -0.8 &  337 & Galaxy\\ 
05D1hn & 02:24:36.25 & -04:10:54.9 & 0.1489\,$\pm$\,0.0006 &  -7.2\,$\pm$\,0.1 & 1.06\,$\pm$\,0.02 &  -0.8 &  232 & Galaxy\\ 
05D1if & 02:24:29.71 & -04:34:13.0 & 0.763\,$\pm$\,0.001 &  -5.0\,$\pm$\,0.3 & 1.03\,$\pm$\,0.03 &  -0.0 & 5000 & Galaxy\\ 
05D1ix & 02:24:19.95 & -04:40:11.7 & 0.49\,$\pm$\,0.01 &  -8.7\,$\pm$\,0.2 & 1.05\,$\pm$\,0.01 &  -0.7 & 5000 & Template Fit\\ 
05D1iy & 02:27:39.97 & -04:25:21.3 & 0.2478\,$\pm$\,0.0003 &  -9.2\,$\pm$\,0.1 & 0.88\,$\pm$\,0.01 &  -0.7 &   52 & Galaxy\\ 
05D2le & 10:01:54.85 & +02:05:34.7 & 0.7002\,$\pm$\,0.0005 &  +6.3\,$\pm$\,0.4 & 1.10\,$\pm$\,0.04 &  -0.0 & 2647 & Galaxy\\ 
05D2mp & 09:59:08.61 & +02:12:14.6 & 0.3537\,$\pm$\,0.0004 &  -3.9\,$\pm$\,0.2 & 1.11\,$\pm$\,0.02 &  +0.7 &  831 & Galaxy\\ 
\enddata
\tablenotetext{a}{The SNLS ID format is explained in \citet{2005ApJ...634.1190H}; briefly, it relates to
the year of discovery (03, 04 etc.), the relevant SNLS field (D1
through D4), and a running ID (e.g. aa, ab, ac etc.).}
\tablenotetext{b}{``Effective'' rest-frame phase relative to maximum light i.e. $(T_{\mathrm{obs}}-T_{\mathrm{max}})/(s\times(1+z))$}
\tablenotetext{c}{Estimated through the spectral extraction aperture; increases greater than 5000\% are listed as 5000\%.}
\tablenotetext{d}{Light curve deemed inadequate in temporal coverage.}
\tablenotetext{e}{Spectrum deemed inadequate in signal-to-noise ratio.}
\tablenotetext{f}{Spectroscopically peculiar; excluded from primary analysis sample.}

\label{tab:sninfo}
\end{deluxetable}

\clearpage

\begin{deluxetable}{lccccccc}
\tabletypesize{\small}
\tablecaption{Observing log}
\tablehead{\colhead{SN Name} & \colhead{Date Obs} & \colhead{MJD} & \colhead{Seeing} & \colhead{Slit} & \colhead{Dichroic} & \colhead{Exptime}& \colhead{SN apparent}\\ \colhead{} & \colhead{} & \colhead{} & \colhead{(\arcsec)} & \colhead{(\arcsec)} & \colhead{} & \colhead{(s)} & \colhead{i' mag.}}
\startdata
03D1au & 2003-09-23 & 52905.51 & 0.8 & 1.0 & 560 &  5400 & 22.4\\ 
03D1aw & 2003-09-23 & 52905.58 & 0.8 & 1.0 & 560 &  5400 & 22.7\\ 
03D1co & 2003-11-22 & 52965.48 & 0.9 & 1.0 & 680 &  9000 & 23.5\\ 
03D1dj & 2003-11-21 & 52964.47 & 0.8 & 1.0 & 560 &  6000 & 22.0\\ 
03D3af & 2003-04-07 & 52736.41 & 0.9 & 1.0 & 680 &  5400 & 22.7\\ 
03D3aw & 2003-05-06 & 52765.53 & 0.9 & 1.0 & 680 &  4800 & 22.2\\ 
03D3ay & 2003-05-06 & 52765.47 & 0.9 & 1.0 & 680 &  3600 & 21.7\\ 
03D3ba & 2003-05-06 & 52765.38 & 0.9 & 1.0 & 680 &  3600 & 21.7\\ 
03D3bb & 2003-05-06 & 52765.29 & 0.9 & 1.0 & 560 &  2000 & 19.8\\ 
03D3bh & 2003-05-06 & 52765.32 & 0.9 & 1.0 & 560 &  3600 & 20.9\\ 
03D3bl & 2003-06-01 & 52791.33 & 0.9 & 1.0 & 680 &  3600 & 22.0\\ 
03D3cc & 2003-06-02 & 52792.27 & 1.0 & 1.0 & 680 &  5400 & 22.1\\ 
03D3cd & 2003-06-01 & 52791.27 & 1.0 & 1.0 & 680 &  9777 & 22.2\\ 
03D4ag & 2003-07-02 & 52822.54 & 0.9 & 1.0 & 560 &  5400 & 21.1\\ 
03D4cj & 2003-08-27 & 52878.32 & 0.9 & 1.0 & 560 &  4000 & 21.2\\ 
03D4dh & 2003-09-23 & 52905.43 & 0.8 & 1.0 & 560 &  3600 & 22.6\\ 
03D4gl & 2003-11-22 & 52965.26 & 0.9 & 1.0 & 680 &  7200 & 22.6\\ 
04D1hd & 2004-09-21 & 53269.47 & 0.9 & 1.0 & 560 &  6000 & 21.6\\ 
04D1jg & 2004-09-21 & 53269.56 & 0.9 & 1.0 & 560 &  6000 & 22.5\\ 
04D1oh & 2004-10-19 & 53297.54 & 1.0 & 1.0 & 560 &  7200 & 22.9\\ 
04D1rh & 2004-12-14 & 53353.24 & 0.9 & 0.7 & 560 &  6000 & 21.9\\ 
04D1sk & 2004-12-14 & 53353.36 & 0.9 & 1.0 & 560 &  9000 & 23.2\\ 
04D2gc & 2004-04-22 & 53117.00 & 0.7 & 0.7 & 560 &  6600 & 22.5\\ 
04D2kr & 2004-12-14 & 53353.58 & 0.9 & 1.0 & 560 &  9600 & 22.8\\ 
04D3cp & 2004-04-23 & 53118.45 & 0.8 & 0.7 & 560 &  2700 & 23.1\\ 
04D3ez & 2004-04-22 & 53117.00 & 0.8 & 0.7 & 560 &  3600 & 21.0\\ 
04D3fk & 2004-04-22 & 53117.00 & 0.8 & 0.7 & 560 &  4500 & 22.3\\ 
04D4in & 2004-09-21 & 53269.31 & 0.9 & 1.0 & 560 &  7200 & 22.4\\ 
04D4jr & 2004-10-19 & 53297.28 & 1.0 & 1.0 & 560 &  7200 & 22.0\\ 
05D1hk & 2005-11-30 & 53704.37 & 0.9 & 1.0 & 560 &  2400 & 21.5\\ 
05D1hn & 2005-11-30 & 53704.40 & 0.9 & 1.0 & 560 &  2100 & 20.8\\ 
05D1if & 2005-12-01 & 53705.31 & 0.9 & 1.0 & 560 &  9000 & 23.2\\ 
05D1ix & 2005-11-30 & 53704.28 & 0.9 & 1.0 & 560 &  7200 & 22.9\\ 
05D1iy & 2005-11-30 & 53704.43 & 0.9 & 1.0 & 560 &  4800 & 22.1\\ 
05D2le & 2005-12-01 & 53705.57 & 0.9 & 1.0 & 560 &  7200 & 23.2\\ 
05D2mp & 2005-11-30 & 53704.53 & 0.9 & 1.0 & 560 &  6600 & 21.9\\ 
\enddata
\label{tab:observinglog}
\end{deluxetable}

\clearpage

\begin{deluxetable}{cccc}
\tablecaption{Mean SN~Ia spectrum (Early)}
\tablehead{\colhead{Wavelength (\AA)} & \colhead{Flux} & \colhead{Flux lower} & \colhead{Flux upper}}
\startdata
2805 & 0.506 & 0.302 & 0.629\\
2810 & 0.579 & 0.322 & 0.643\\
2815 & 0.417 & 0.364 & 0.662\\
2820 & 0.477 & 0.390 & 0.682\\
2825 & 0.482 & 0.431 & 0.708\\
2830 & 0.529 & 0.449 & 0.724\\
2835 & 0.747 & 0.472 & 0.757\\
2840 & 0.723 & 0.483 & 0.792\\
2845 & 0.676 & 0.481 & 0.813\\
2850 & 0.719 & 0.495 & 0.830\\
\enddata
\tablecomments{The complete version of this table is in the electronic edition of
the Journal.  The printed edition contains only a sample.}
\label{tab:mean_spectra_early}
\end{deluxetable}

\newpage
\begin{deluxetable}{cccc}
\tablecaption{Mean SN~Ia spectrum (Max)}
\tablehead{\colhead{Wavelength (\AA)} & \colhead{Flux} & \colhead{Flux lower} & \colhead{Flux upper}}
\startdata
2805 & 0.287 & 0.262 & 0.411\\
2810 & 0.360 & 0.282 & 0.422\\
2815 & 0.372 & 0.299 & 0.435\\
2820 & 0.438 & 0.317 & 0.448\\
2825 & 0.355 & 0.334 & 0.456\\
2830 & 0.477 & 0.349 & 0.470\\
2835 & 0.433 & 0.357 & 0.472\\
2840 & 0.378 & 0.367 & 0.483\\
2845 & 0.442 & 0.375 & 0.496\\
2850 & 0.428 & 0.377 & 0.506\\
\enddata
\tablecomments{The complete version of this table is in the electronic edition of
the Journal.  The printed edition contains only a sample.}
\label{tab:mean_spectra_max}
\end{deluxetable}

\clearpage

\end{document}